\documentclass[letterpaper]{article}

\usepackage[T1]{fontenc}

\usepackage{geometry}
\usepackage{setspace}

\usepackage[
backend=biber,
style=chem-acs,
articletitle=true,
maxnames=99,
autocite=superscript,
sorting=none,
sortcites=true
]{biblatex}
\usepackage{graphicx}
\usepackage{float}
\newfloat{scheme}{htbp}{los}
\floatname{scheme}{Scheme}
\floatname{chart}{Chart}
\newfloat{graph}{htbp}{loh}

\usepackage{chemformula} 
\usepackage[version = 4]{mhchem} 

\usepackage{amssymb, amsmath, amsthm}
\usepackage{algorithm, algpseudocode}   

\usepackage{cleveref} 
\usepackage{graphicx}
\usepackage{listings}
\usepackage{epstopdf}
\usepackage{multirow}
\usepackage{booktabs}
\usepackage{xcolor}
\usepackage{url}
\usepackage{subcaption}
\usepackage[USenglish]{babel}

\newcommand{\etal}{\textit{et al. }}

\usepackage{authblk}
\author[1]{Hua Huang*}
\affil[1]{NVIDIA, Santa Clara, USA 94086}
\author[2]{Wenkai Shao}
\affil[2]{Individual researcher, Anqing, China 246000}
\author[3]{Jeff Hammond}
\affil[3]{NVIDIA, Helsinki, Finland 00180}

\title{Accelerating Density Fitting with Adaptive-precision and 8-bit Integer on AI Accelerators}
\date{*Email: huah@nvidia.com}

\begin{document}

\maketitle

\begin{abstract}
The emergence of artificial intelligence (AI) accelerators like NVIDIA Tensor Cores offers new opportunities to speed up tensor-heavy scientific computations. However, applying them to quantum chemistry is challenging due to strict accuracy demands and irregular data patterns. We propose an adaptive precision algorithm to accelerate the density fitting (DF) method with Gaussian basis sets on AI accelerators using 8-bit integer (INT8) arithmetics. Implemented in the GPU-accelerated PySCF package, the algorithm is tested on more than twenty molecular systems with different NVIDIA GPUs. Compared to the standard FP64 code, our algorithm is up to 204\% faster on a RTX 4090 gaming GPU and up to 364\% faster on a RTX 6000 Ada workstation GPU without compromising the converged energy. This work demonstrates a practical approach to use AI hardware for reliable quantum chemistry simulations.
\end{abstract}

\section*{Keywords}

density fitting, mixed-precision, matrix multiplication, AI accelerator



\section{Introduction}

The recent surge in artificial intelligence (AI) applications has driven the development of hardware accelerators with extremely high throughput for low-precision matrix operations. Notably, NVIDIA Tensor Cores\supercite{V100whitepaper,A100whitepaper,H100whitepaper} and Google TPUs\supercite{TPU:2017} offer order-of-magnitude improvements in dense general matrix multiplication (GEMM) performance using half- and single-precision and even lower precision arithmetic. While these accelerators have transformed deep learning, their impact on scientific computing is limited. One major barrier is that scientists typically consider double precision (FP64), which is also supported on Tensor Cores on some GPUs, as the standard for scientific computing. Moreover, although GEMM is a common computational primitive in both AI and scientific computing, it does not encompass the spectrum of mathematical operations needed in scientific computing.

To address these challenges, researchers have proposed a variety of single- and mixed-precision algorithms for scientific computing. These include general purpose methods, such as mixed-precision solvers for systems of linear equations\supercite{Abdelfattah:2021,Higham:2022,Kashi:2025}, as well as techniques that emulate a higher precision GEMM using multiple lower precision GEMMs\supercite{Ozaki:2012,Mukunoki:2020,Ootomo:2022,Ootomo:2024,Ozaki:2025}. In addition, domain-specific low- and mixed-precision algorithms have also been developed. For instance, molecular dynamics simulations often perform force calculations in single precision while retaining double precision for critical steps like integration and energy accumulation. However, due to the complexity and irregularity of their computational patterns, most of these algorithms cannot fully exploit the GEMM units designed for AI. Given that GEMM units can deliver low-precision performance that is several times, or even an order of magnitude, higher than that of double precision, developing tailored algorithms that effectively leverage them could result in substantial performance gains.

Researchers have proposed different mixed-precision algorithms to accelerate quantum chemistry computations with different types of basis. For the plane wave basis, a very recent work by Hou \etal proposed using single precision in Hartree-Fock exchange computation with low-rank approximation techniques\supercite{Hou:2025}. For real-space basis sets, Das \etal employed single precision for some matrix multiplications in Cholesky QR decomposition and the Rayleigh-Ritz procedure in the DFT-FE package\supercite{Das:2019}. For Gaussian basis, several early works\supercite{Ufimtsev:2008,Asadchev:2010,Miao:2013} and a recent study by Laqua \etal\supercite{Laqua:2021} investigated computing four-center and three-center electron repulsion integrals (ERIs) on CUDA cores using mixed double and single precision. However, none of these methods leverage specialized GEMM units. Habib \etal proposed a mixed-precision matrix factorization of the inverse overlap matrix with Tensor Cores\supercite{Habib:2024}, but did not analyze the proportion of matrix decomposition time in one iteration or the speed up of the entire DFT calculation using this method.

In this paper, we present an adaptive precision density fitting (DF) approach for accelerating DFT computation with Gaussian basis on Tensor Cores with 8-bit signed integer (INT8) GEMMs. The contraction of large tensors forms the bulk of the computation in DF\supercite{Whitten:1973,Eichkorn:1995,Weigend:2002,Sodt:2006,Aquilante:2007,Aquilante:2009,Reine:2008,Merlot:2013}, making it particularly well suited for acceleration using Tensor Cores. We propose an adaptive precision strategy for DF that converges to the same energies, using the same number or no more than two extra self-consistent field (SCF) iterations. Our method is implemented and validated in the GPU-enabled PySCF package\supercite{Sun:2018,Sun:2020,Li:2024,Wu:2024}. It can be readily integrated into other quantum chemistry, offering a promising pathway to accelerate quantum chemical simulations without compromising accuracy. Our contributions are summarized as follows:
\begin{itemize}
\item We propose an adaptive precision algorithm utilizing INT8 GEMM on Tensor Core for DF calculation.
\item Our implementation in PySCF achieves up to 364\% full-DFT speedup on NVIDIA RTX 6000 Ada, confirming the practical benefits of precision-aware Tensor Core usage.
\item We demonstrate robust convergence behavior and numerical stability across a wide range of molecules and basis sets of our method.
\end{itemize}

\section{Background}

\subsection{Density Fitting}
\label{sect:bg:df}

Density fitting (DF), also called resolution-of-the-identity (RI), is a method to approximate the ERI tensor. DF uses a set of $N_{aux}$ auxiliary basis functions $\{\hat{\psi}_p\}$ to fit the ERI tensor:
\begin{equation}
\label{equ:df}
(ij|lk) \approx \sum_{p,q} (ij|p) \widehat{J}^{-1}_{pq} (q|kl).
\end{equation}
In practice, we compute and store
\begin{equation}
\label{equ:cderi}
B^{q}_{ij} = \sum_{p} (ij|p) L_{pq}^{-1},
\end{equation}
where $LL^{\mathsf{T}} = \widehat{J}$ is the Cholesky decomposition of $\widehat{J}$. Using $B^{q}_{ij}$ and the density matrix $D$, the Coulomb matrix $J$ is computed in two steps:
\begin{align}
\label{equ:df-j-1}
Y^q &= \sum_{ij} B^q_{ij} D_{ij}, \\
\label{equ:df-j-2}
J_{ij} &= \sum_q B^q_{ij} Y^q.
\end{align}
The exchange matrix $K$ is also computed in two steps. Let $C_{rs}$ denote the expansion coefficients for the $s$-th occupied molecular orbital. Then, assuming symmetry of $D$, we compute 
\begin{align}
\label{equ:df-k-1}
W^q_{is} &= \sum_{r} B^q_{ir} C_{rs}, \\
\label{equ:df-k-2}
K_{ij} &= \sum_{q} \sum_{s} W^q_{is} W^q_{js}.
\end{align}

Schwarz screening is applied in density fitting to reduce the cost of certain computations\supercite{Werner:2003}. Shell pair $MN$ will be neglected if 
\begin{equation*}
\sqrt{(MN|MN) \text{max}_{P} (P|P)} < \tau,
\end{equation*}
where $\max_P (P|P)$ is the maximum value of $(P|P)$ on all auxiliary basis shells. The sparsity pattern of $B^q_{ij}$ is the same as the sparsity pattern of $(ij|p)$ after screening, since the contraction in Formula \ref{equ:cderi} is in the dimension of the auxiliary basis set.

\subsection{Single- and Mixed-Precision Algorithms for Quantum Chemistry}

The rapid development of general-purpose GPUs (GPGPUs) in the past 15 years has prompted research on how to utilize the computing, especially in single precision, capabilities of GPGPUs to accelerate quantum chemical calculations. In this section, we briefly review the efforts to accelerate quantum chemistry calculations with single- and mixed-precision.

Multiple single precision and mixed-precision algorithms have been proposed for Gaussian basis sets. One of the earliest works by Ufimtsev \etal\supercite{Ufimtsev:2008} proposed computing ERIs on GPU using mixed double- and single precision. Depending on domain-specific limits, ``important'' ERIs are calculated in double precision while others are calculated in single precision. The authors claim that this approach gives a nearly $2\times$ speed-up without compromising the accuracy as a double precision calculation. However, these early algorithms for computing ERI on GPUs can only handle smaller basis sets and are therefore less practical. More recently, Laqua \etal also proposed computing ``significant'' three-center-one-electron (3c1e) integrals using double precision and less ``significant'' 3c1e integrals in single precision, and these 3c1e integrals are used exclusively in the semi-numerical integration method\supercite{Laqua:2021} developed by the same authors. Performance improvement and convergence behaviors in this work are comparable to the work of Ufimtsev \etal 

For real space basis, Das \etal contributed a DFT package DFT-FE using finite element basis with mixed-precision acceleration\supercite{Das:2019}. The key part of DFT-FE is a Chebyshev polynomial filtered subspace iteration (CheFSI)\supercite{Zhou:2006} for solving electron orbital eigenvalues and eigenvectors. In their implementation, multiple GEMMs within the Cholesky QR decomposition and Rayleigh-Ritz procedures were carefully selected for single precision calculations. However, even for an extremely large system with 6164 Mg atoms, single precision GEMMs take about 35\% of the overall running time. In most practical cases, those single precision GEMMs will provide only a very limited improvement in performance.  Dawson \etal\supercite{Dawson2024RedPrecisionQC} demonstrated the use of mixed precision in a density matrix purification solver, including both FP32 and FP16 steps to achieve FP64 results.

For DFT using the plane wave basis, a very recent work by Hou \etal\supercite{Hou:2025} proposed using single precision in the interpolative separable density fitting (ISDF)\supercite{Lu:2015,Hu:2017-ISDF,Dong:2018} and the adaptively compressed exchange (ACE)\supercite{Lin:2016,Hu:2017-ACE} operator. Fast Fourier transforms (FFTs), the computational bottleneck, are also accelerated using single precision. This approach achieves a nearly $2\times$ speedup for multiple test cases.

Multiple works have also investigated using single precision for post-HF calculations. Olivares-Amaya \etal explored the use of single- and mixed-precision GEMM for tensor contractions in the second-order Møller-Plesset perturbation theory (MP2)\supercite{Olivares:2010}. DePrince \etal studied using single precision GEMMs for tensor contractions in coupled cluster theory with single and double excitations (CCSD)\supercite{DePrince:2011}. They also suggested doing one or more final iterations in double precision after the iterations in single precision to reduce the error of the converged energy to the required level. The work of Pokhilko \etal further suggests that using single precision for correlation energy calculation is sufficient for many-body methods in a wide range of practical applications\supercite{Pokhilko:2018}.

The newly released cuEST\supercite{cuEST:2026} library supports emulated FP64 GEMM in density fitting, which was developed independently from the current effort. A user can use the default emulation level in cuEST or manually specify one emulation level.  cuEST also supports the ``float float'' approximation\supercite{Thall2006ExtendedPrecision} for FP64 in the VV10 functional\supercite{Vydrov2010JCP} evaluation, which is computationally intensive but does not involve matrix multiplication.

\subsection{Floating-point Arithmetic and Low-precision Emulation}

A floating-point value consists of three parts: a sign bit, an exponent part, and a mantissa (fraction) part. Then a normalized value $x$ in a floating-point value set is represented as follows:
\begin{equation}
x = (-1)^s \times 1.m_1m_2\cdots m_{(l_m - 1)} \times 2^e,
\end{equation}
where $s$ is the sign bit, $e$ is the exponent part (a signed integer), and $1.m_1m_2\cdots m_{(l_m - 1)}$ is the mantissa part in binary number, $l_m$ is the number of bits of the mantissa part. The first bit in the mantissa part is not explicitly stored. Different floating point data types have different numbers of bits of both the mantissa part and the exponent part, for example, FP64 has 11 bits exponent and 52 (+1 hidden) bits mantissa, and single precision (FP32) has 8 bits exponent and 23+1 bits mantissa.

Floating-point emulation methods use multiple low-precision floating-point or integer numbers to represent one high-precision floating-point number and use multiple low-precision floating-point or integer arithmetic operations to simulate one high-precision floating-point arithmetic operations. In this work, we focus on emulating FP64 dense general matrix multiplication (GEMM) using low-precision GEMMs.

The Ozaki scheme\supercite{Ozaki:2012,Ozaki:2013} is one of the earliest floating-point GEMM emulation schemes that can achieve any accuracy. Mukunoki \etal proposed a method to compute the Ozaki scheme on Tensor Cores using half-precision (FP16, 5 bits exponent and 10+1 bits mantissa) GEMMs\supercite{Mukunoki:2020}. This FP16 emulation approach represents one FP64 value $x^{\text{fp64}}$ as the scaled sum of multiple FP16 values (``splits'') $\hat{x}^{\text{fp16}}_i$:
\begin{equation}
x^{\text{fp64}} = \sum_{i=1}^s \left( \hat{x}^{\text{fp16}}_i \times 2^{-e_i} \right),
\end{equation}
where $e_i$ are positive integers, and the FP64 GEMM is emulated by $s(s+1)/2$ FP16 GEMMs. One major disadvantage of this approach is the mantissa waste bits in FP16. The work of Ootomo \etal\supercite{Ootomo:2024} pointed out that each FP16 split can only capture $\alpha < 11$ bits of mantissa from the original FP64 input. The ``bits per one split'' (BPS) value $\alpha$ is computed as 
\begin{equation}
\alpha = \min(11, \lfloor (24 - \log_2k) / 2 \rfloor),
\end{equation}
where $k$ is the size of the inner product of matrix multiplication. 
In other words, for a larger $k$ value, the FP16 emulation approach needs to increase $s$ to maintain accuracy, which leads to a rapid decrease in performance. It is possible to split the original GEMM in the inner product dimension to reduce the $k$ value in each GEMM and avoid wasting too many mantissa bits, but a smaller problem size also harms the GEMM performance.

In the past two years, new research has focused on emulating FP64 GEMM using 8-bit signed integer (INT8) GEMMs. Ootomo \etal first employed the Ozaki scheme with INT8 GEMMs\supercite{Ootomo:2024} and their implementation ozIMMU is released on GitHub\supercite{Ootomo:ozIMMU}. ozIMMU handles the exponent part and the mantissa part separately and directly emulate the addition and multiplication between mantissa parts using INT8. Ozaki \etal further proposed a new method\supercite{Ozaki:2025} called Ozaki scheme II, which scales and converts FP64 inputs to INT8 using Chinese Remainder Theorem, then performs INT8 GEMMs. Their implementation GEMMul8 is also released on GitHub\supercite{Ozaki:GEMMul8}. These INT8 emulation algorithms significantly outperform the FP16 emulation algorithm.

\section{Accelerate DF with Adaptive Precision and INT8 GEMMs on Tensor Cores}

We implemented an adaptive precision scheme that utilizes INT8-emulated GEMMs to accelerate DF calculations. In summary, the algorithm is:
\begin{itemize}
\item Compute the Coulomb matrix $J$ using Formulas (\ref{equ:df-j-1}) and (\ref{equ:df-j-2}) in FP64;
\item Compute the Hartree-Fock exchange matrix $K$ using Formulas (\ref{equ:df-k-1}) and (\ref{equ:df-k-2}) and INT8-emulated FP64 GEMM with adaptive precision determined by the relative change magnitude of SCF energy;
\item Fall back to direct FP64 GEMM when SCF is almost converge and INT8-emulated FP64 GEMM is no longer faster than direct FP64 GEMM since the cost of emulated FP64 GEMM increases with the emulation accuracy.
\end{itemize}
The details of the algorithm will soon be discussed in Section \ref{sect:mxp-alg}.

\subsection{Justification of Accelerating only DF $K$ Matrix Build}

We choose DF instead of the direct method because the computations in DF are better suitable for Tensor Cores. Although tensor contractions are also one of the performance bottlenecks in the direct methods, the sizes of contracted tensors in the direct method are very small. Let $M$, $N$, $P$, $Q$ denote shell indices. Then blocks of the Coulomb and exchange matrices are
\begin{equation}
J_{MN} = \sum_{PQ} (MN|PQ) D_{PQ} \text{\quad and\quad} K_{MN} = \sum_{PQ} (MP|NQ) D_{PQ}, 
\end{equation}
respectively, where $(MN|PQ)$ denotes a shell quartet of ERI. The sizes of each $(MN|PQ)$ tensor depend on the selected basis and the shell indices $M$, $N$, $P$, and $Q$; in most cases the size of $(MN|PQ)$ smaller than $21^4$. These tensor contractions are computed as matrix-vector multiplication after reordering and fusion of tensor indices, while the matrix is smaller than $441 \times 441$ in most cases. Such matrix-vector multiplications with small matrices are not suitable for Tensor Cores. Furthermore, exploiting the 8-way symmetry property leads to irregular memory access patterns to $(MN|PQ)$ tensor and $J$, $K$, $D$ matrices. Some earlier work by Huang \etal\supercite{Huang:2018,Huang:2020} has discussed these challenges and proposed several strategies to mitigate them on CPUs, but these techniques cannot be applied to Tensor Cores. In contrast, density fitting only involves four large tensor contractions, which are much more suitable for GPUs and Tensor Cores.

The second question is why only the exchange matrix $K$ is computed using the adaptive precision approach. The first and most important reason is that building the $K$ matrix requires significantly more floating point operations (FLOPs) and has a much longer computation time compared to building the $J$ matrix. The second reason is that the constructions of the $J$ and $V$ matrices cannot be expressed as GEMMs. We now proceed to estimate the FLOP counts for constructing both the $J$ and $K$ matrices.

After applying Schwarz screening to the $(ij|$ pairs of the 3c2e integrals $(ij|p)$, only a subset of the $(ij|$ pairs survives. Let $N_{scr}$ denote the number of pairs $(ij|$ that remain after screening. $N_{scr} \le N_{bf}^2$ always holds, and normally $N_{scr} = \Theta(N_{bf})$ for large $N_{bf}$. For each pair of indexes $(q, s)$, the computation in Formula (\ref{equ:df-k-1}) is to multiply a sparse matrix $B_{ir}$ with a dense vector $C_r$. We can then compute the number of FLOPs required in each step:
\begin{itemize}
\item Formulas (\ref{equ:df-j-1}) and (\ref{equ:df-j-2}): $2 N_{scr} N_{aux}$,
\item Formula (\ref{equ:df-k-1}): $2 N_{scr} N_{aux} N_{occ}$,
\item Formula (\ref{equ:df-k-2}): $2 N_{bf}^2 N_{aux} N_{occ}$.
\end{itemize}
Thus, the $J$ matrix build requires a total of $4 N_{scr} N_{aux}$ FLOPs, while the $K$ matrix build requires $2 (N_{scr} + N_{bf}^2) N_{aux} N_{occ}$ FLOPs, which is at least $N_{occ}$ times more than the FLOPs needed for the $J$ matrix build. This significant difference in computational cost motivates us to keep the $J$ matrix computation in FP64, while accelerating the $K$ matrix computation using the adaptive precision scheme.

\subsection{Adaptive Precision DF $K$ Matrix Build with INT8 GEMM}
\label{sect:mxp-alg}

We use the newly released emulated FP64 GEMM in cuBLAS 13.0 update 2 for computing the $K$ matrix. The emulated FP64 GEMM implementation in cuBLAS uses a fixed point emulation and follows the Ozaki scheme, and it
allows the user to set the number of mantissa bits $m^{\text{emu}}$ to be used for fixed emulation\supercite{cuBLASdoc}. The implementation uses $s^{\text{emu}} = \lceil (m^{\text{emu}} + 1) / 8 \rceil$ splits of INT8 to represent the input. For details about this algorithm, we recommend that readers refer to a recent work of Schwarz \etal\supercite{Schwarz:2026}.

Consequently, we use an \textit{adaptive precision selection} scheme in the SCF iterations. Since the SCF procedure is iterative, different initial guesses of the solution can often converge to the same final result with the same or slightly different numbers of iterations. If the early iterations are computed with reduced precision, it can be considered as using a slightly different initial guess, and the overall convergence may remain unchanged. Let $E^{\text{total}}_i$ denote the total energy after the $i$-th SCF iteration, and let 
\begin{equation}
\Delta E_{i} = E^{\text{total}}_i - E^{\text{total}}_{i-1}, \quad 
\Delta E_{i}^{\text{rel}} = |\Delta E_{i} / E^{\text{total}}_i|.
\end{equation}
$|\Delta E_{i}|$ is usually used as a criterion to determine if the SCF iteration has converged. We base our precision selection on $\Delta E_{i}^{\text{rel}}$, the relative change in total energy between two adjacent SCF iterations, since relative errors are more meaningful in numerical analysis and methods. We select $m^{\text{emu}}_{i+1}$ for the $(i+1)$-th SCF iteration based on Table \ref{table:mxp-scheme} and hardware specifications. For example, we return to the standard FP64 calculation for $\Delta E^{\text{rel}}_i < 2 \times 10^{-6}$ (emulation level <= 3) on H100 GPU since the H100 Tensor Cores support FP64 and the standard FP64 calculation is faster than the INT8 emulation using 10 or more INT8 GEMMs. 

\begin{table}[htb]
\centering
\begin{tabular}{ccccc}
\hline
$\Delta E^{\text{rel}}_i$ & \begin{tabular}[c]{@{}c@{}}Emu.\\ Level\end{tabular} & $m^{\text{emu}}_{i+1}$ & \begin{tabular}[c]{@{}c@{}}Emu. Relative\\ Accuracy\end{tabular} & \begin{tabular}[c]{@{}c@{}}\# INT8\\ Splits\end{tabular}  \\
\hline
$\ge 2 \times 10^{-6}$                       & 4 & 23     & $1.2 \times 10^{-7}$    & 3    \\
$\in [5 \times 10^{-9}, 2 \times 10^{-6})$   & 3 & 31     & $4.7 \times 10^{-10}$   & 4    \\
$\in [2 \times 10^{-11}, 5 \times 10^{-9})$  & 2 & 39     & $1.8 \times 10^{-12}$   & 5    \\
$\in [7 \times 10^{-14}, 2 \times 10^{-11})$ & 1 & 47     & $7.1 \times 10^{-15}$   & 6    \\
$< 7 \times 10^{-14}$                        & 0 & (FP64) & ($2.2 \times 10^{-16}$) & (0)  \\ 
\hline
\end{tabular}
\caption{Adaptive precision selection scheme based on SCF total energy convergence}
\label{table:mxp-scheme}
\end{table}

The algorithm starts with low precision ($m^{\text{emu}}_{1} = 23$) and gradually transitions to higher precision as the total energy approaches convergence. As SCF iterations converge, $\Delta E_{i+1}$ could be 10 times smaller than $\Delta E_i$ if the computation is performed in FP64. Meanwhile, $m^{\text{emu}}_{i+1}$ used in the $(i+1)$-th SCF iteration is determined by $\Delta E^{\text{rel}}_i$ instead of $\Delta E^{\text{rel}}_{i+1}$. Therefore, we set $\Delta E^{\text{rel}}_i$ thresholds to promote emulation accuracy approximately 10 times higher than the current relative accuracy of INT8 emulation. In addition, $|\Delta E_i|$ in standard FP64 SCF calculations does not always decrease monotonically. To stabilize the computation, we disallow reverting to a lower precision once a higher precision has been used. 

\subsection{Implementation Details}
\label{sect:impl-detail}

We implement our adaptive precision DF algorithm in the PySCF package\supercite{Sun:2018}, which leverages the CuPy library\supercite{Okuta:2017} for GPU acceleration. 

The original PySCF GPU implementation computes and stores the 3D tensor $B^q_{ij}$ in a symmetric sparse form (only non-zero $B^q_{ij}$ values with $i \ge j$ are stored) to reduce memory usage. Then, it partitions the computations along the $q$ dimension. Using MATLAB colon notation, the computation of Formulas (\ref{equ:df-k-1}) and (\ref{equ:df-k-2}) in PySCF can be rewritten as:
\begin{gather}
\label{equ:df-k-blk-1}
W^{q1:q2}_{is} = \sum_{r} B^{q1:q2}_{ir} C_{rs}, \\
\label{equ:df-k-blk-2}
K_{ij} = K_{ij} + \sum_{q=q1}^{q2} \sum_{s} W^q_{is} W^q_{js}.
\end{gather}
Each tensor block $B^{q1:q2}_{ir}$ is first symmetrized and restored to a dense tensor, then $W^{q1:q2}_{is}$ is calculated using the dense tensor. In addition, $\sum_{q=q1}^{q2} \sum_{s} W^q_{is} W^q_{js}$ is computed using one GEMM instead of one symmetric rank-k update (SYRK). We follow these practices in our implementation for three reasons. First, leveraging sparsity requires repacking the corresponding rows or columns of $C(r, s)$ to match the sparsity pattern of $B^q_{ij}$, which introduces nontrivial data movement and additional memory overhead. Second, the original tensor contraction needs to be computed as many GEMMs with different sizes of matrices, but these sizes are often not large enough to fully saturate the GPU in a single call. One potential optimization is to use grouped GEMM kernels from the CUTLASS library\supercite{CUTLASS:2023}, which support batched GEMMs with varying input dimensions. However, this feature is currently only available on NVIDIA H100 and newer Blackwell GPUs. Considering these factors, we opt not to utilize the sparsity in $B^q_{ij}$. Lastly, the INT8 emulation code is also not optimized for SYRK yet. Therefore, we follow the practices in the PySCF GPU implementation. Future implementations in C/C++-based quantum chemistry libraries may explore sparsity-aware and symmetry-aware optimizations more effectively.

To enable DFT calculations with density fitting on GPUs for larger molecular systems, we adjust several GPU memory usage thresholds in PySCF. In particular, we allow up to 80\% of the available GPU memory to be used to store the sparse $B^q_{ij}$. We pad the dimensions of $B^{q1:q2}_{ir}$ and $C_{rs}$ to multipliers of 32, allowing cuBLAS to select and use GEMM kernels optimized for Tensor Cores.

\section{Numerical Experiments}

\subsection{Experiment Setup}

We evaluate our mixed-precision density fitting algorithm across  multiple basis sets and molecular systems on a range of NVIDIA GPUs. Specifically, we use three types of GPUs in our experiments: a consumer-grade RTX 4090, a workstation-class RTX A6000, and a server-grade flagship H100. \Cref{table:gpu-specs} summarizes their key specifications. These GPUs are installed on different host systems; however, all systems share the same operating system and software environment, detailed as follows:
\begin{itemize}
\item Ubuntu 24.04 LTS with kernel 6.8.0;
\item NVIDIA GPU driver 580.105.8 with CUDA toolkit 13.1;
\item Python 3.12 + PySCF 2.8.0 + cupy-cuda13x 13.6.0 + cutensor-cu13 2.3.1.
\end{itemize}

We employ two high-accuracy basis sets in our tests: the split-valence Pople basis set 6-311G(d,p)\supercite{Krishnan:1980} and the Karlsruhe basis set def2-TZVPP\supercite{Weigend:2005,Weigend:2006}.  The corresponding auxiliary basis sets are cc-pVTZ-JKFIT and def2-TZVPP-JKFIT, respectively. For all DFT calculations, we use the widely adopted hybrid exchange-correlation functional B3LYP\supercite{Becke:1993,Stephens:1994}. 

The test set includes diverse molecular systems representing different dimensionalities and complexities: 1D alkane chains, 2D graphene-like molecules, 3D water clusters, and 3D organic molecules. \Cref{table:molecular-systems} provides a comprehensive summary of the tested molecular systems, their associated basis sets, and reference DFT results computed using the standard FP64 implementation. All reference results are obtained with DF on a single NVIDIA H100 GPU with 80 GB of HBM3 memory.

\begin{table}[htb]
\centering
\begin{tabular}{cccc}
\hline
                    & RTX 4090     & RTX 6000 Ada & H100         \\
\hline
Architecture        & \multicolumn{2}{c}{Ada Lovelance} & Hopper \\
Compute Capability  & \multicolumn{2}{c}{8.9}  & 9.0             \\
Mem. Size (GB)      & 24           & 48        & 80              \\
Mem. Bandwidth      & 1008 GB/s    & 960 GB/s  & 3.35 TB/s       \\
\# CUDA cores       & 16384        & 18176     & 16896           \\
\# Tensor Cores     & 512          & 568       & 528             \\
Peak FP64 (TFLOPS)  & 1.3          & 1.4       & 33.5/67$^{*}$   \\
Peak FP32 (TFLOPS)  & 82.6         & 91.1      & 67              \\
Peak TC INT8 (TOPS) & 660.6$^{*}$  & 728.5$^{*}$ & 1979$^{*}$    \\
\hline
\end{tabular}
\caption{Key specifications of the GPUs used in numerical experiments. The asterisk indicates that the Tensor Cores of this GPU support GEMM of this data type, and the corresponding values are the performance of Tensor Cores.}
\label{table:gpu-specs}
\end{table}

\begin{table}[htb]
\centering
\begin{tabular}{ccrrrrr}
\hline
Molecule & Atoms & $N_{bf}$ & $N_{aux}$ & $N_{occ}$ & \# SCF Iters. & Converged Energy \\
\hline
Alkane-62     & C20H42       & 612  & 2840 & 81  & 9  &  -787.67278936 \\
              &              & 1208 & 2256 & 81  & 9  &  -787.76535664 \\
\hline
Alkane-122    & C40H82       & 1212 & 5620 & 161 & 9  & -1574.13589317 \\
              &              & 2388 & 4476 & 161 & 9  & -1574.32029172 \\
\hline
Alkane-182    & C60H122      & 1812 & 8400 & 241 & 9  & -2360.59899665 \\
              &              & 3568 & 6696 & 241 & 9  & -2360.87522643 \\
\hline
Graphene-36   & C24H12       & 504  & 2256 & 78  & 9  &  -922.07474766 \\
              &              & 912  & 2016 & 78  & 9  &  -922.19791221 \\
\hline
Graphene-72   & C54H18       & 1080 & 4806 & 171 & 10 & -2069.31091985 \\
              &              & 1926 & 4374 & 171 & 10 & -2069.58764198 \\
\hline
(H2O)20       & (H2O)20      & 600  & 2780 & 100 & 12 & -1529.16143540 \\
              &              & 1180 & 2260 & 100 & 12 & -1529.43152890 \\
\hline
(H2O)47       & (H2O)47      & 1410 & 6533 & 235 & 10 & -3593.63712813 \\
              &              & 2773 & 5311 & 235 & 11 & -3594.23830243 \\
\hline
Tamoxifen     & C26H29N1O1   & 678  & 3082 & 100 & 11 & -1138.22247481 \\
              &              & 1274 & 2626 & 100 & 11 & -1138.37146192 \\
\hline
Sphingomyelin & C25H51N2O5P1 & 908  & 4179 & 135 & 13 & -1810.65324984  \\
              &              & 1748 & 3460 & 135 & 13 & -1810.90326016  \\
\hline
\end{tabular}
\caption{Molecular systems used for numerical experiments and reference converged energy. For each molecule, the first and the second rows correspond to 6-311G(d,p) and def2-TZVPP basis sets, respectively. $N_{bf}$ and $N_{aux}$ are the numbers of basis functions of the original and DF auxiliary basis sets, respectively.  $N_{occ}$ is the number of occupied orbitals. The reference converged energy are computed using standard FP64 calculations with DF.}
\label{table:molecular-systems}
\end{table}

\subsection{Single SCF Iteration Performance Improvement}
\label{sect:single-scf-perf}

\begin{figure}[htb]
\begin{subfigure}{0.32\textwidth}
  \centering
  \caption{RTX 4090, Alkane-122/6-311G(d,p)}
  \includegraphics[width=\linewidth]{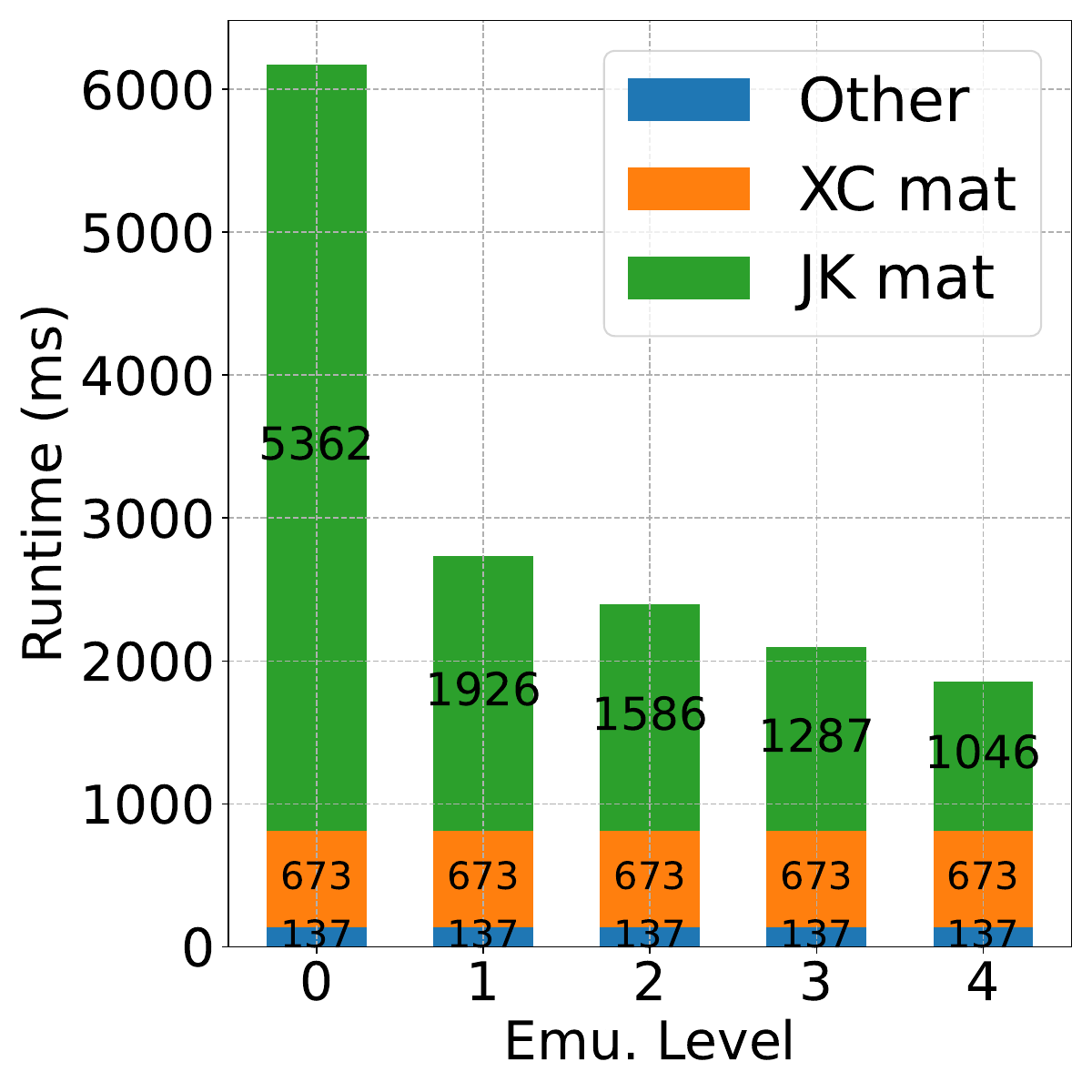}  
\end{subfigure}
\begin{subfigure}{0.32\textwidth}
  \centering
  \caption{RTX 6000 Ada, Alkane-182/6-311G(d,p)}
  \includegraphics[width=\linewidth]{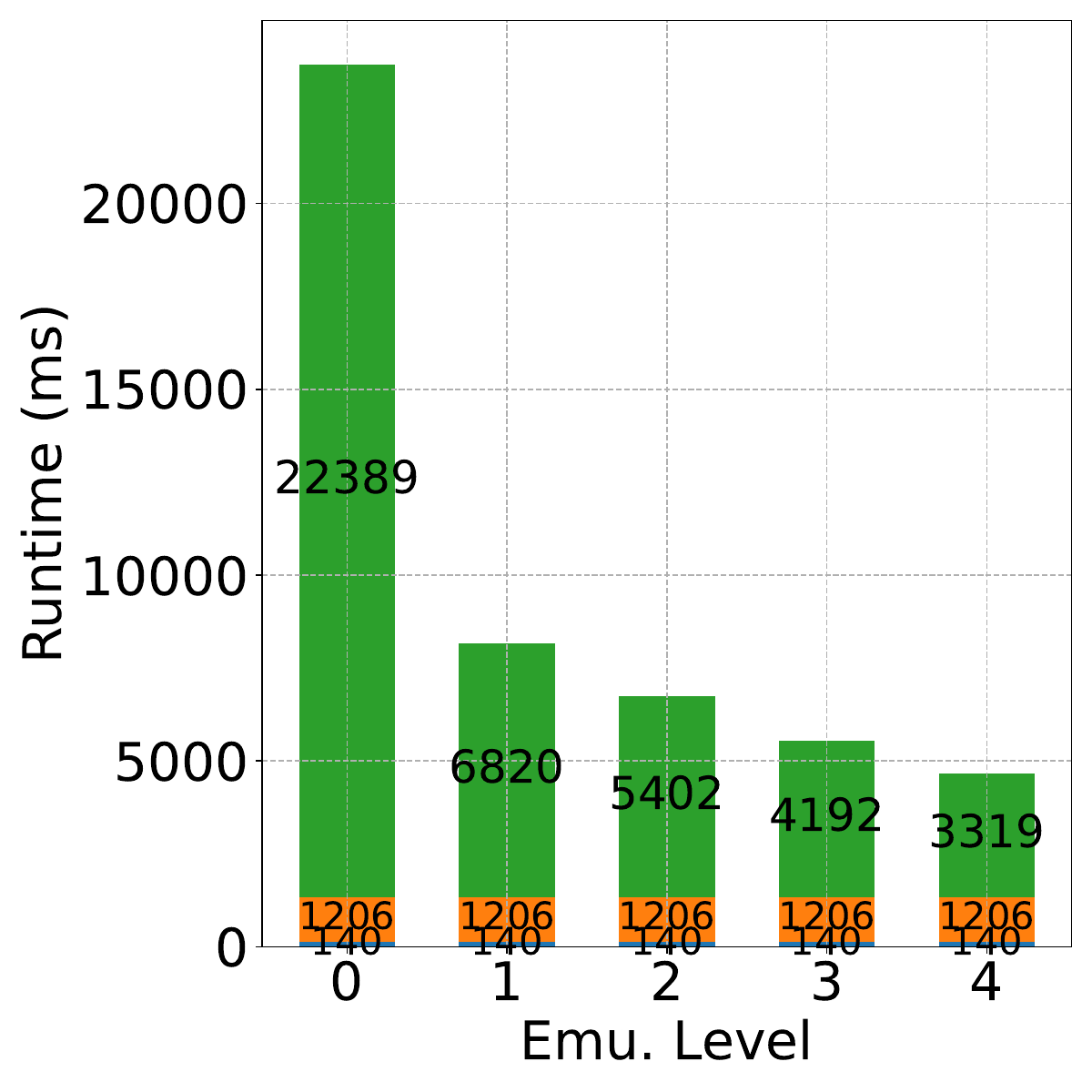}
\end{subfigure}
\begin{subfigure}{0.32\textwidth}
  \centering
  \caption{H100, Alkane-182/def2-TZVPP}
  \includegraphics[width=\linewidth]{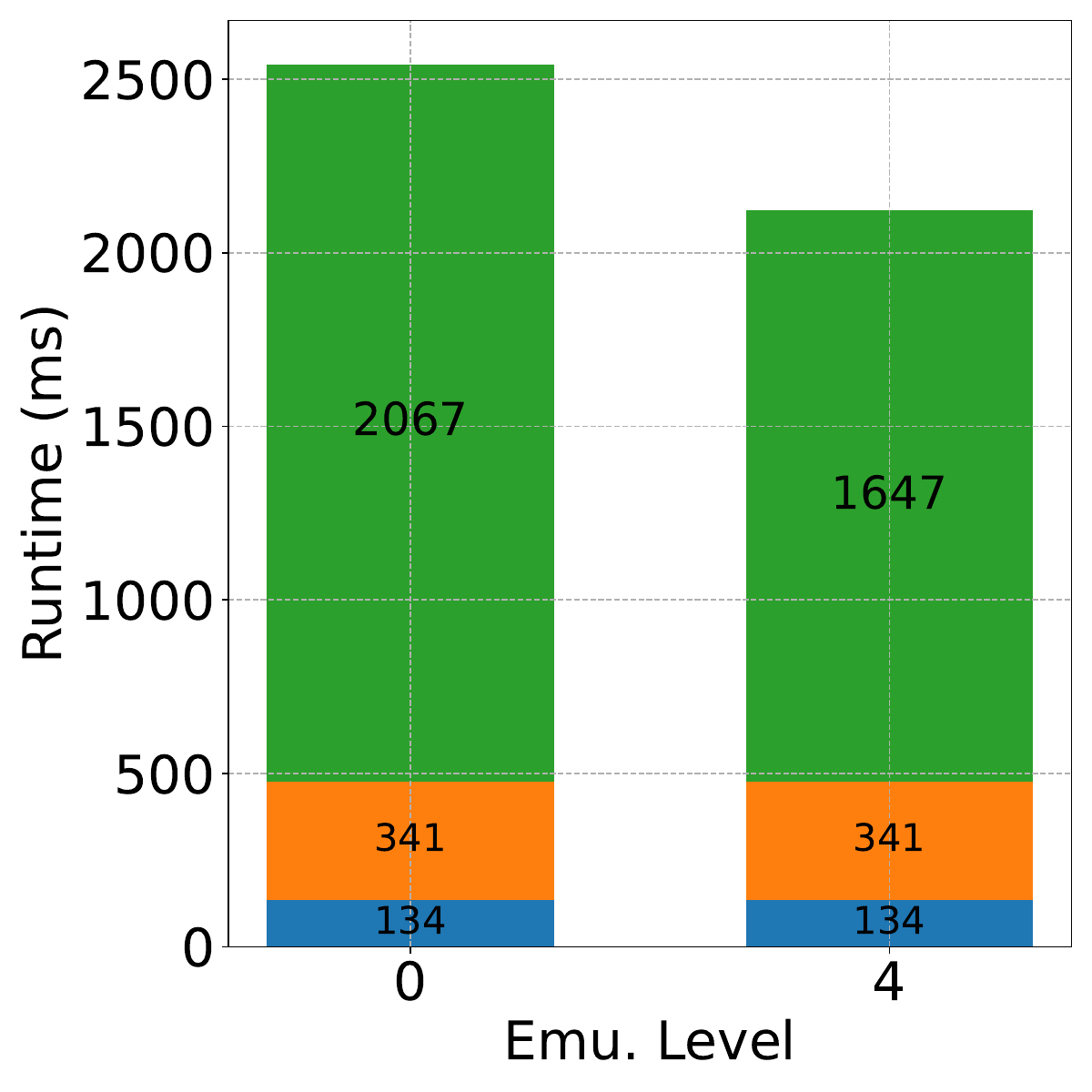}
\end{subfigure}
\caption{Running time breakdowns (in milliseconds) of the second SCF iteration on different GPUs and molecular systems: (a) Alkane-122/6-311G(d,p) on a RTX 4090 (left); (b) Alkane-182/6-311G(d,p) on a RTX 6000 Ada (middle); and (c) Alkane-182/def2-TZVPP on a H100 (right).}
\label{fig:runtime-breakdown}
\end{figure}

We first study the performance improvement of a single SCF iteration. Figure \ref{fig:runtime-breakdown} shows the running time breakdowns (in milliseconds) of the second SCF iteration on different GPUs. For each GPU, we choose the largest molecular system it can run in the test set. The time for constructing the $J$ matrix is not listed separately since it is too small (smaller than 50 ms in all three test cases). For all molecular systems and GPUs tested, the construction of the $K$ matrix is the most time-consuming part. On RTX 4090 and RTX 6000 Ada, switching from FP64 to INT8 emulation significantly accelerates the construction of the $K$ matrix, since these GPUs have weak FP64 vector performance and no FP64 support on their Tensor Cores. In addition, the performance of different levels of INT8 emulation is different: level 4 (3xINT8 splits) only needs about half the time of level 1 (6xINT8 splits). On H100, only level 4 INT8 emulation is slightly faster than the standard FP64 implementation, since H100 Tensor Cores support FP64. These results also show the necessary of the proposed adaptive precision algorithm instead of simply using INT8 emulated FP64 GEMM as a drop-in replacement of the native FP64 GEMM.

\subsection{DFT Calculation Convergence}
\label{sect:dft-convergence}

\begin{table}[htb]
\centering
\begin{tabular}{c|cc|cc}
\hline
              & \multicolumn{2}{c|}{6-311G(d,p)} & \multicolumn{2}{c}{def2-TZVPP} \\ 
              \cline{2-5} 
              & Standard      & Adaptive         & Standard     & Adaptive        \\
              & FP64          & Precision        & FP64         & Precision       \\
\hline
Alkane-62     & 9   & {\it 10} / {\it 10} / 9 & 9   & {\it 10} / {\it 10} / 9   \\
Alkane-122    & 9   & {\it 10} / {\it 10} / 9 & 9   & - / {\it 10} / 9          \\
Alkane-182    & 9   & - / {\it 10} / 9        & 9   & 9 / - / -                 \\
Graphene-36   & 9   & 9 / 9 / 9               & 9   & 9 / 9 / 9                 \\
Graphene-72   & 10  & 10 / {\it 11} / 10      & 10  & 10 / - / -                \\
(H2O)20       & 12  & 12 / 12 / 12            & 12  & 12 / 12 / 12              \\
(H2O)47       & 10  & - / {\it 11} / 10       & 11  & 11 / - / -                \\
Tamoxifen     & 11  & 11 / 11 / 11            & 11  & 11 / 11 / 11              \\
Sphingomyelin & 13  & 13 / 13 / 13            & 13  & 13 / 13 / -               \\
\hline
\end{tabular}
\caption{Number of SCF iterations needed for convergence on different GPUs. Values in {\it italic} indicate a higher iteration count than that of the standard FP64 algorithm. For adaptive precision, the reported values are the numbers on RTX 4090 / RTX 6000 Ada / H100, respectively. A short dash indicates the GPU does not have enough memory to run this test case.}
\label{table:convergece}
\end{table}

In this section, we evaluate the convergence of the proposed adaptive precision algorithm. 

Table \ref{table:convergece} reports the number of SCF iterations required to reach $|\Delta E| \le 10^{-7}$ on different GPUs. We note that different quantum chemistry packages use different default thresholds: $10^{-9}$ in PySCF, $10^{-7}$ in NWChem\supercite{NWChem:2020}, and $10^{-6}$ in both Psi4\supercite{Psi4:2020} and ORCA\supercite{ORCA:2020}. We checked all converged energies of the adaptive precision algorithm; the absolute errors between these energies and the reference converged energies computed by the standard FP64 algorithm are less than $10^{-7}$. For RTX 4090 and RTX 6000 Ada, the adaptive precision algorithm uses the same number of SCF iterations as the standard FP64 algorithm to converge; a few systems require one extra SCF iteration. For H100, the adaptive precision algorithm has the same convergence behavior as the standard FP64 algorithm. Since the adaptive precision algorithm returns to FP64 on H100 when $\Delta E^{\text{rel}}_i < 2 \times 10^{-6}$ instead of $\Delta E^{\text{rel}}_i < 7 \times 10^{-14}$, the convergence of H100 is slightly better. 

\begin{figure}[]
\centering
\includegraphics[width=\linewidth]{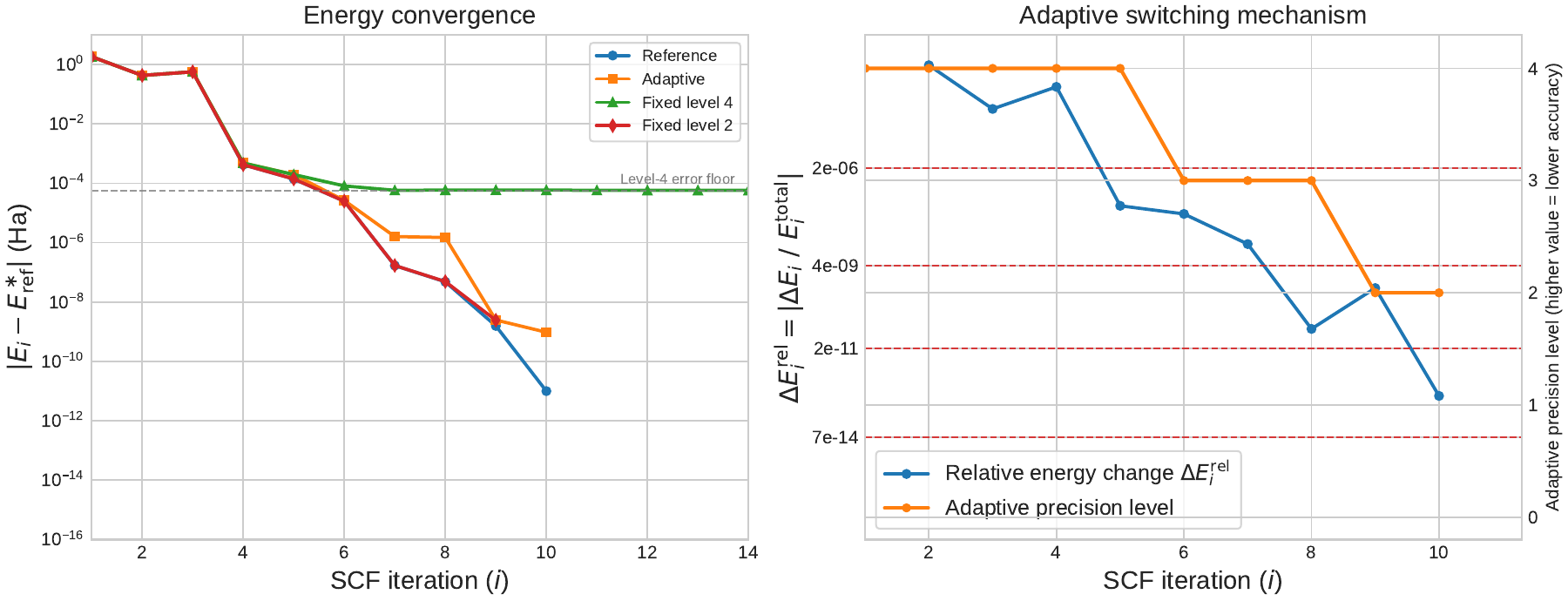}
\caption{SCF energy convergence with different mixed-precision schemes on an RTX 6000 Ada GPU for molecular system Alkane-122 / 6-311G(d,p) (left) and the adaptive precision switching mechanism (right).}
\label{fig:energy-conv-adaptive-vs-fixed}
\end{figure}

\Cref{fig:energy-conv-adaptive-vs-fixed} demonstrates both the effectiveness of the adaptive mixed-precision scheme and the rationale behind its switching thresholds. In the left panel, we plot the SCF energy error relative to the converged reference value as a function of SCF iteration. All methods show similar early-stage error reduction, but the fixed level-4 run reaches a clear error floor at approximately $10^{-5}$ Ha, after which additional iterations provide little improvement. By contrast, the adaptive schedule continues to reduce the error to the $10^{-9}$ Ha range, close to the high-accuracy reference result. The right panel shows that the adaptive precision level decreases from 4 to 2 as the relative energy change $\Delta E_{i}^{\text{rel}}$ decreases. This threshold-driven schedule avoids the low-precision error floor while retaining reduced precision in the early SCF regime, where it is most efficient.

To summary, if a molecular system has good convergence using the standard FP64 DF algorithm, our proposed adaptive precision algorithm should also have good convergence. 

\subsection{DFT Calculation Performance Improvement}
\label{sect:dft-perf}

\begin{figure}[]
\centering
\includegraphics[width=\linewidth]{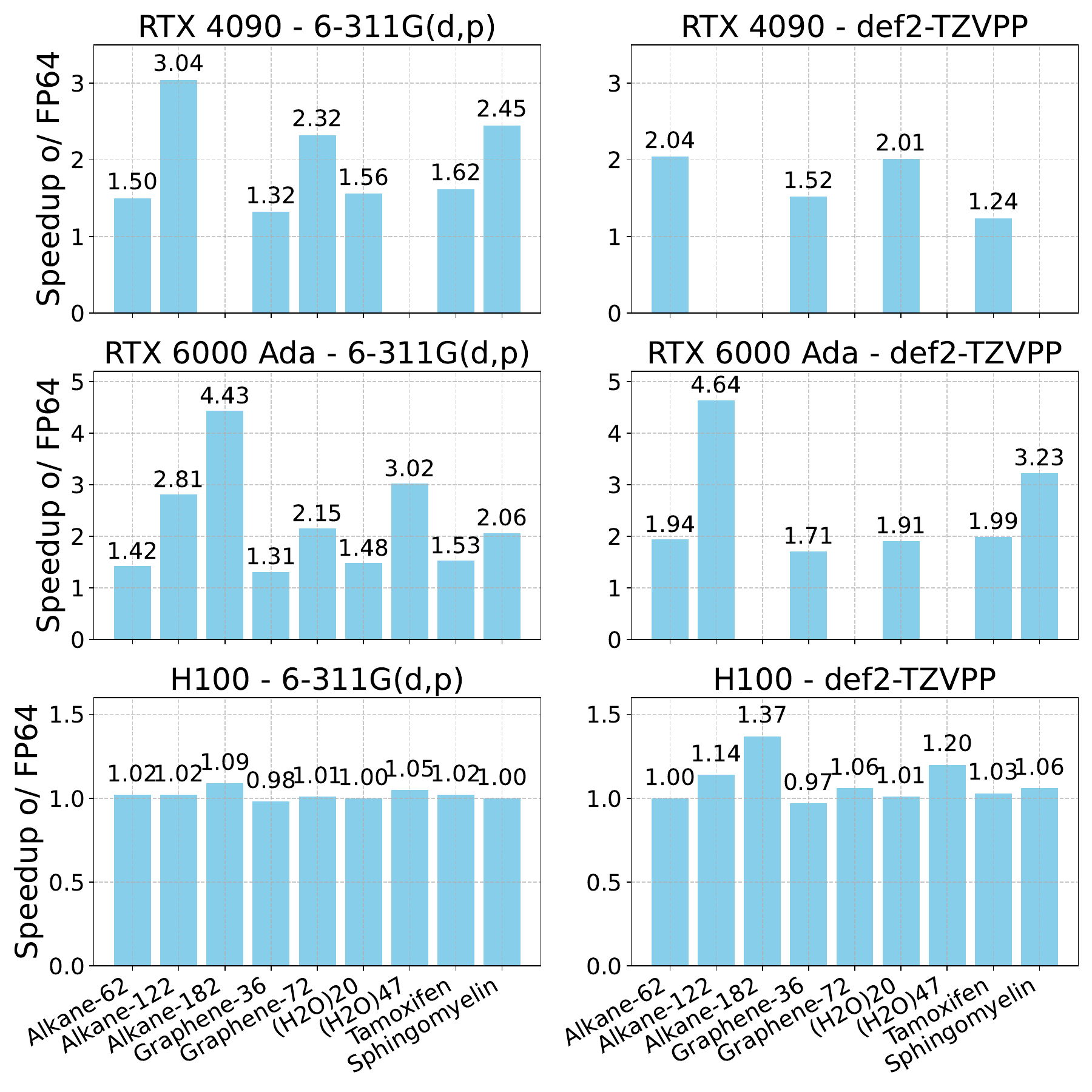}  
\caption{Speedup of DFT calculations when using the adaptive precision method over using the standard FP64 algorithm. All computational steps, including the integrals $(ij|p)$ and the construction of the $B_{ij}^q$ tensor in Formula~(\ref{equ:cderi}), are included. Missing data points indicate that the GPU did not have sufficient memory to complete the corresponding calculation.}
\label{fig:dft-speedup}
\end{figure}

Finally, we analyze the overall performance improvement of a DFT calculation using adaptive precision algorithms. All computational steps are included, including the integrals $(ij|p)$ and the construction of the $B_{ij}^q$ tensor in Formula~(\ref{equ:cderi}). \Cref{fig:dft-speedup} presents the speedup of using the adaptive precision algorithm over the standard FP64 algorithm. 

On all GPUs, the adaptive precision algorithm has a larger speedup over the standard FP64 algorithm on a larger molecular system, since the time required to construct the $K$ matrix accounts for a larger proportion of the total running time of a larger molecular system. On RTX 4090 and RTX 6000 Ada, the adaptive precision algorithm has a speedup of up to 3.04 and 4.64, respectively. For molecular systems that the adaptive precision algorithm needs one more SCF iteration to converge, for example, Alkane-122/6-311G(d,p), the adaptive precision algorithm still has a very large speedup. These two GPUs have the same architecture and similar numbers of Tensor Cores, while RTX 6000 Ada has twice the memory size of RTX 4090. Therefore, RTX 6000 Ada can run larger systems and have a larger acceleration on larger systems. On H100, even if its Tensor Cores are designed to accelerate both traditional scientific workloads and modern AI tasks with FP64 and low-precision data types, the adaptive precision algorithm can still reach a speedup of 1.37. Even if the overhead of INT8 emulation cancels out the performance improvement of adaptive precision algorithm on small molecular systems, only 3\% overall performance loss at most on a H100. To summary, using the adaptive precision algorithm is beneficial in most cases.

\begin{figure}[]
\centering
\includegraphics[width=\linewidth]{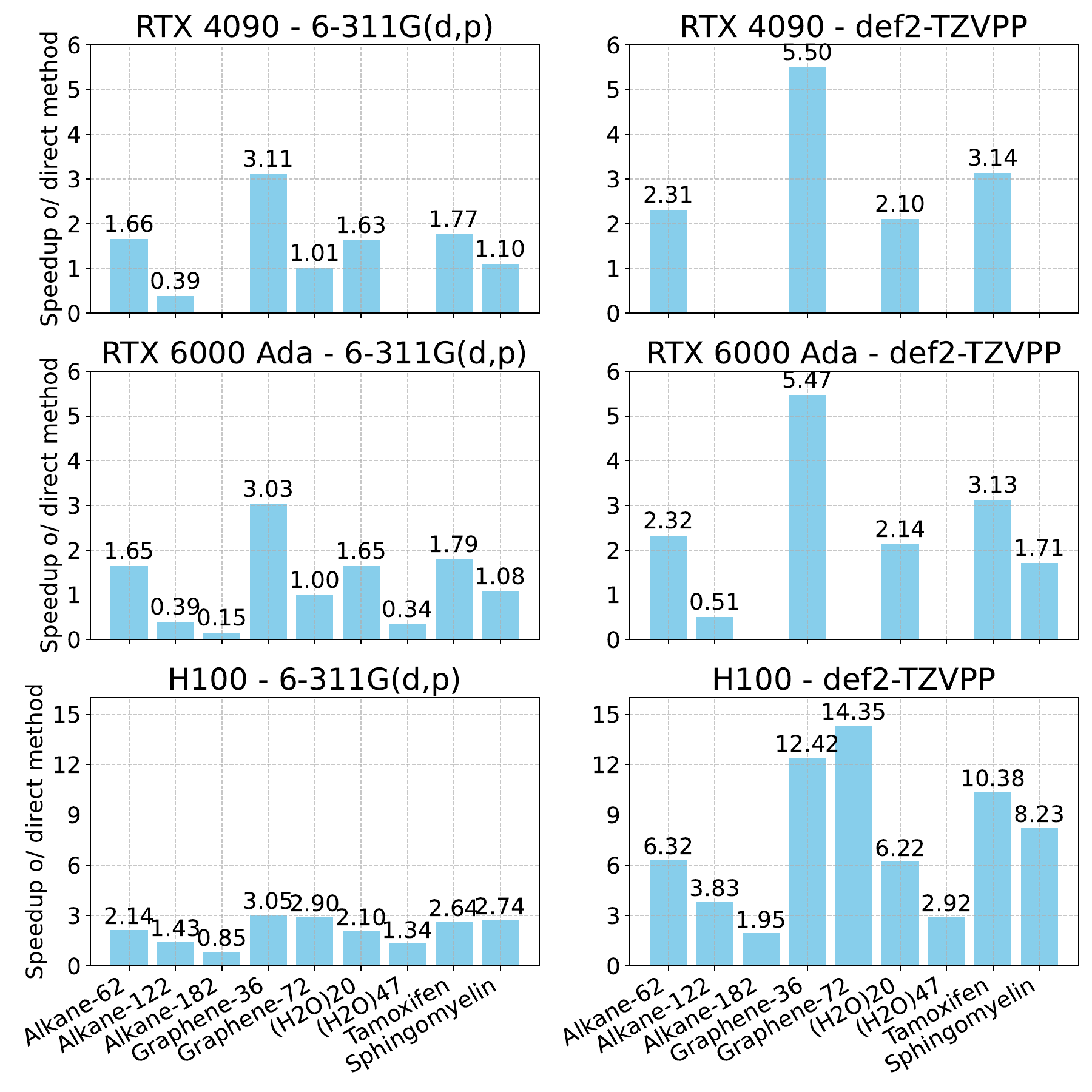}  
\caption{Speedup of DFT calculations when using FP64 density fitting over using the FP64 direct method. Missing data points indicate that the GPU did not have sufficient memory to complete the corresponding DF calculation.}
\label{fig:df-fp64-speedup}
\end{figure}

\begin{figure}[]
\centering
\includegraphics[width=\linewidth]{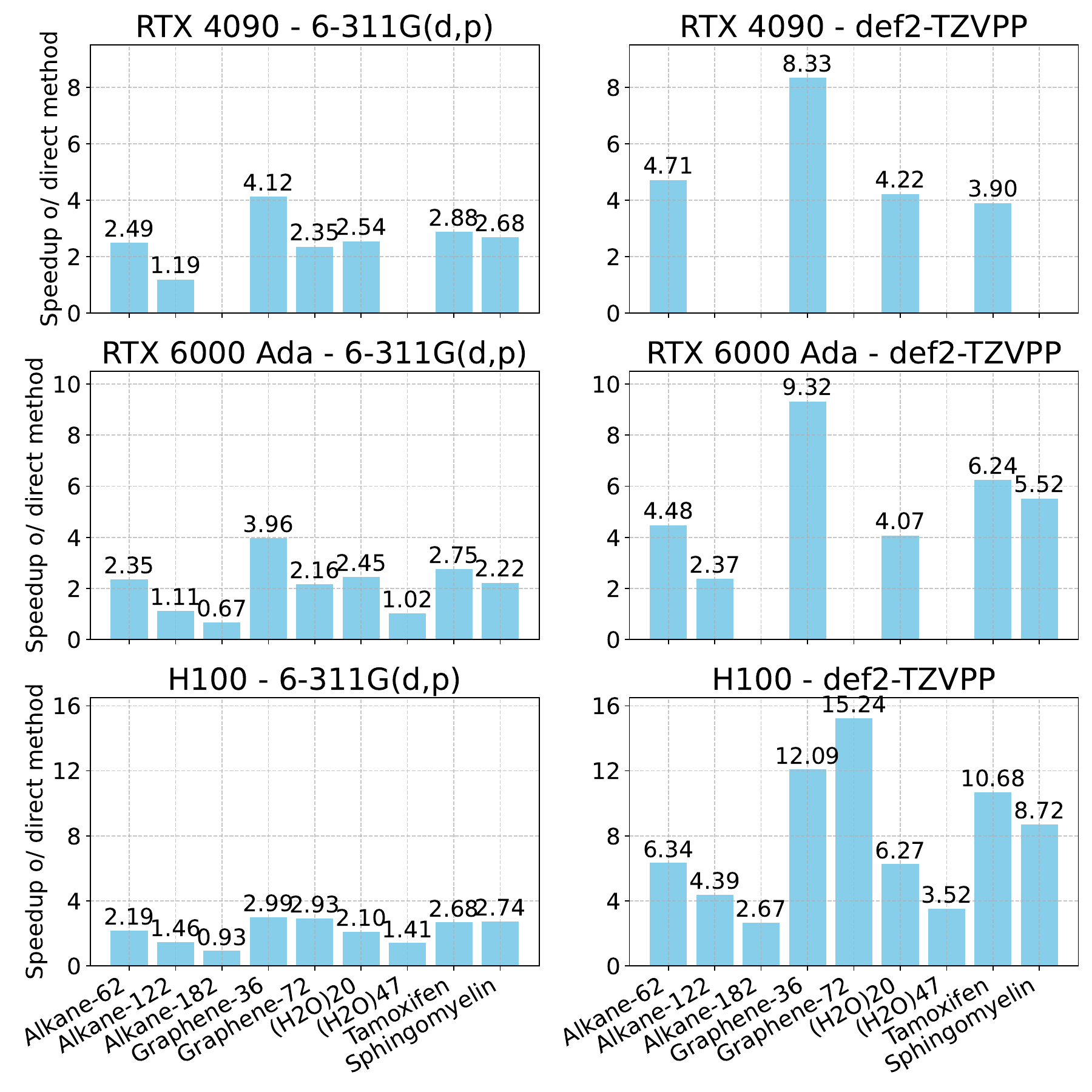}  
\caption{Speedup of DFT calculations when using density fitting with adaptive precision method over using the FP64 direct method. Missing data points indicate that the GPU did not have sufficient memory to complete the corresponding DF calculation.}
\label{fig:df-mxp-speedup}
\end{figure}

We also compared the performance between density fitting and the direct method on the same GPUs. \Cref{fig:df-fp64-speedup} and \Cref{fig:df-mxp-speedup} present the speedup of using the standard FP64 density fitting method over the FP64 direct method and the speedup of using the adaptive precision density fitting method over the FP64 direct method, respectively. The results show that density fitting method with and without adaptive precision has a large advantage over the direct method for most test cases. For Alkane-122 and Alkane-182, direct method is faster than density fitting method as the direct method can better leverage the sparsity created by screening on long alkane molecules.

\subsection{Implementing the Adaptive Precision Algorithm with Other Libraries}

We also implemented our adaptive precision algorithm with the ozIMMU library\supercite{Ootomo:ozIMMU} and the GEMMul8 library\supercite{Ozaki:GEMMul8} and tested their performance. One common and significant issue of our ozIMMU-based and GEMMul8-based implementations is the inaccurate estimation of the emulation level to achieve desired accuracy levels. Unlike cuBLAS, ozIMMU and GEMMul8 do not support emulating FP64 GEMM to a specific accuracy level. We have to estimate the number of INT8 slices to use in ozIMMU and GEMMul8 based on the results in their paper and adjust \Cref{table:mxp-scheme}, and we might slightly over-estimate the number of slices in some cases. As a result, our ozIMMU-based and GEMMul8-based implementations can still achieve a good speedup on some large systems (for example, 6-311G(d,p)+Alkane-182) compared to the standard FP64 GEMM implementation, but no speedup on moderate or small size systems (detailed results omitted). These results also show the advantage of the new emulated FP64 GEMM APIs in cuBLAS. 
\section{Conclusion and Future Work}

In this work, we propose an adaptive precision algorithm to accelerate the construction of exchange matrix in the widely used density fitting method. Our algorithm exploits the newly proposed INT8 emulation of FP64 GEMM and the property of iterative methods to fully utilize computational capacities of AI accelerators. We implemented our algorithm in the GPU version of PySCF for evaluation. Numerical experiment results show that our algorithm can deliver up to 204\%, 365\%, and 37\% speedup on a gaming GPU, a workstation GPU, and a flagship GPU while maintaining a high accuracy of DFT calculation.

Experiment results also point out some issues that need further research. The first topic is numerical stability and error consideration. Although the numerical stability of INT8 emulation has been studied in earlier works, our strategies of switching between emulation accuracies are empirical and have no formal error bounds or convergence guarantees. We admit that it is very hard to perform formal numerical error analysis for complicated non-linear iterative algorithms. In the future, we will try to further refine our algorithm by linking precision and emulation level to residual norms or density matrix changes.

The second topic is the use of single precision (FP32) to construct the exchange-correlation (XC) matrix. Figure \cref{fig:runtime-breakdown} shows that once the $K$ matrix construction is significantly accelerated, building the XC matrix could be a new performance bottleneck. PySCF and many other quantum chemistry packages use the LibXC library\supercite{Lehtola:2018} to evaluate the values of different XC functionals on a set of points for constructing the $V$ matrix. LibXC supports FP32, but it needs to be compiled separately. This means that the quantum chemistry library will need to handle linking and invoking two LibXC libraries. In addition, the selection of XC functional evaluation points also has a large impact on the accuracy of DFT. When using FP32, the number of evaluation points should be reduced compared to using FP64 to achieve a balance between accuracy and computational cost.

\section*{Acknowledgements}

The authors thank Minseok Lee for the discussion of utilizing Tensor Cores 
in scientific computing and Hiroyuki Ootomo for the meaningful discussion 
of FP64 GEMM emulation using low-precision GEMM operations.



\printbibliography

@misc{V100whitepaper,
	title        = {{NVIDIA} {Tesla} {V100} {GPU} Architecture},
	author       = {NVIDIA},
	year         = 2017,
	urldate      = {2025-04-04},
	howpublished = {https://images.nvidia.com/content/volta-architecture/pdf/volta-architecture-whitepaper.pdf}
}

@misc{A100whitepaper,
	title        = {{NVIDIA} {A100} {Tensor Core} {GPU} Architecture},
	author       = {NVIDIA},
	year         = 2020,
	urldate      = {2025-04-04},
	howpublished = {https://images.nvidia.com/aem-dam/en-zz/Solutions/data-center/nvidia-ampere-architecture-whitepaper.pdf}
}

@misc{H100whitepaper,
	title        = {{NVIDIA} {H100} {Tensor Core} {GPU} Architecture},
	author       = {NVIDIA},
	year         = 2022,
	urldate      = {2025-04-04},
	howpublished = {https://resources.nvidia.com/en-us-tensor-core}
}

@misc{cuBLASdoc,
	title        = {{cuBLAS} Release 13.2 Documentation},
	author       = {NVIDIA},
	year         = 2026,
	urldate      = {2026-03-16},
	howpublished = {https://docs.nvidia.com/cuda/cublas/}
}

@inproceedings{TPU:2017,
	title        = {In-datacenter performance analysis of a tensor processing unit},
	author       = {Jouppi, Norman P. and Young, Cliff and Patil, Nishant and Patterson, David and Agrawal, Gaurav and Bajwa, Raminder and Bates, Sarah and Bhatia, Suresh and Boden, Nan and Borchers, Al and Boyle, Rick and Cantin, Pierre-luc and Chao, Clifford and Clark, Chris and Coriell, Jeremy and Daley, Mike and Dau, Matt and Dean, Jeffrey and Gelb, Ben and Ghaemmaghami, Tara Vazir and Gottipati, Rajendra and Gulland, William and Hagmann, Robert and Ho, C. Richard and Hogberg, Doug and Hu, John and Hundt, Robert and Hurt, Dan and Ibarz, Julian and Jaffey, Aaron and Jaworski, Alek and Kaplan, Alexander and Khaitan, Harshit and Killebrew, Daniel and Koch, Andy and Kumar, Naveen and Lacy, Steve and Laudon, James and Law, James and Le, Diemthu and Leary, Chris and Liu, Zhuyuan and Lucke, Kyle and Lundin, Alan and MacKean, Gordon and Maggiore, Adriana and Mahony, Maire and Miller, Kieran and Nagarajan, Rahul and Narayanaswami, Ravi and Ni, Ray and Nix, Kathy and Norrie, Thomas and Omernick, Mark and Penukonda, Narayana and Phelps, Andy and Ross, Jonathan and Ross, Matt and Salek, Amir and Samadiani, Emad and Severn, Chris and Sizikov, Gregory and Snelham, Matthew and Souter, Jed and Steinberg, Dan and Swing, Andy and Tan, Mercedes and Thorson, Gregory and Tian, Bo and Toma, Horia and Tuttle, Erick and Vasudevan, Vijay and Walter, Richard and Wang, Walter and Wilcox, Eric and Yoon, Doe Hyun},
	year         = 2017,
	booktitle    = {2017 ACM/IEEE 44th Annual International Symposium on Computer Architecture (ISCA)},
	pages        = {1--12},
	doi          = {10.1145/3079856.3080246},
        publisher    = {ACM},
        address      = {New York, NY, USA},
}

@techreport{Abdelfattah:2021,
	title        = {Advances in Mixed Precision Algorithms: 2021 Edition},
	author       = {Abdelfattah, A. and Anzt, H. and Ayala, A. and Boman, E. and Carson, E. and Cayrols, S. and Cojean, T. and Dongarra, J. and Falgout, R. and Gates, M. and others},
	year         = 2021,
	month        = {08},
	doi          = {10.2172/1814677},
	url          = {https://www.osti.gov/biblio/1814677},
	institution  = {Lawrence Livermore National Lab. (LLNL), Livermore, CA (United States)},
	place        = {United States}
}

@article{Higham:2022,
	title        = {Mixed precision algorithms in numerical linear algebra},
	author       = {Higham, Nicholas J. and Mary, Theo},
	year         = 2022,
	journal      = {Acta Numerica},
	volume       = 31,
	pages        = {347–414},
	doi          = {10.1017/S0962492922000022}
}

@misc{Kashi:2025,
	title        = {Mixed-precision numerics in scientific applications: survey and perspectives},
	author       = {Aditya Kashi and Hao Lu and Wesley Brewer and David Rogers and Michael Matheson and Mallikarjun Shankar and Feiyi Wang},
	year         = 2025,
	url          = {https://arxiv.org/abs/2412.19322},
	eprint       = {2412.19322},
	archiveprefix = {arXiv}
}

@article{Hou:2025,
	title        = {Dual-Grid and Mixed-Precision Methods for Accelerating Plane-Wave Hybrid Functional Electronic Structure Calculations},
	author       = {Hou, Bingkun and Chen, Sheng and Qin, Xinming and Hu, Wei and Yang, Jinlong},
	year         = 2025,
	journal      = {Journal of Chemical Theory and Computation},
	volume       = 21,
	number       = 2,
	pages        = {787--802},
	doi          = {10.1021/acs.jctc.4c01541},
	url          = {https://doi.org/10.1021/acs.jctc.4c01541}
}

@inproceedings{Das:2019,
	title        = {Fast, scalable and accurate finite-element based ab initio calculations using mixed precision computing: 46 {PFLOPS} simulation of a metallic dislocation system},
	author       = {Das, Sambit and Motamarri, Phani and Gavini, Vikram and Turcksin, Bruno and Li, Ying Wai and Leback, Brent},
	year         = 2019,
	booktitle    = {Proceedings of the International Conference for High Performance Computing, Networking, Storage and Analysis},
	location     = {Denver, Colorado},
	publisher    = {ACM},
	address      = {New York, NY, USA},
	series       = {SC '19},
	doi          = {10.1145/3295500.3357157},
	url          = {https://doi.org/10.1145/3295500.3357157},
	articleno    = 2,
	numpages     = 11,
	pages        = {1--11}
}

@article{Miao:2013,
	title        = {Acceleration of Electron Repulsion Integral Evaluation on Graphics Processing Units via Use of Recurrence Relations},
	author       = {Miao, Yipu and Merz, Kenneth M. Jr.},
	year         = 2013,
	journal      = {Journal of Chemical Theory and Computation},
	volume       = 9,
	number       = 2,
	pages        = {965--976},
	doi          = {10.1021/ct300754n}
}

@article{Asadchev:2010,
	title        = {Uncontracted {Rys} Quadrature Implementation of up to {G} Functions on Graphical Processing Units},
	author       = {Asadchev, Andrey and Allada, Veerendra and Felder, Jacob and Bode, Brett M. and Gordon, Mark S. and Windus, Theresa L.},
	year         = 2010,
	journal      = {Journal of Chemical Theory and Computation},
	volume       = 6,
	number       = 3,
	pages        = {696--704},
	doi          = {10.1021/ct9005079}
}

@article{Ufimtsev:2008,
	title        = {Quantum Chemistry on Graphical Processing Units. 1. Strategies for Two-Electron Integral Evaluation},
	author       = {Ufimtsev, Ivan S. and Martínez, Todd J.},
	year         = 2008,
	journal      = {Journal of Chemical Theory and Computation},
	volume       = 4,
	number       = 2,
	pages        = {222--231},
	doi          = {10.1021/ct700268q}
}

@article{Laqua:2021,
	title        = {Accelerating seminumerical {Fock}-exchange calculations using mixed single- and double-precision arithmethic},
	author       = {Laqua, Henryk and Kussmann, Jörg and Ochsenfeld, Christian},
	year         = 2021,
	month        = {06},
	journal      = {The Journal of Chemical Physics},
	volume       = 154,
	number       = 21,
	pages        = 214116,
	doi          = {10.1063/5.0045084}
}

@article{Habib:2024,
	title        = {Efficient Mixed-Precision Matrix Factorization of the Inverse Overlap Matrix in Electronic Structure Calculations with {AI}-Hardware and {GPU}s},
	author       = {Habib, Adela and Finkelstein, Joshua and Niklasson, Anders M. N.},
	year         = 2024,
	journal      = {Journal of Chemical Theory and Computation},
	volume       = 20,
	number       = 16,
	pages        = {7102--7112},
	doi          = {10.1021/acs.jctc.4c00584}
}

@article{Sun:2018,
	title        = {{PySCF}: the {Python}-based simulations of chemistry framework},
	author       = {Sun, Qiming and Berkelbach, Timothy C. and Blunt, Nick S. and Booth, George H. and Guo, Sheng and Li, Zhendong and Liu, Junzi and McClain, James D. and Sayfutyarova, Elvira R. and Sharma, Sandeep and Wouters, Sebastian and Chan, Garnet Kin-Lic},
	year         = 2018,
	journal      = {WIREs Computational Molecular Science},
	volume       = 8,
	number       = 1,
	pages        = {e1340},
	doi          = {10.1002/wcms.1340}
}

@article{Sun:2020,
	title        = {Recent developments in the {PySCF} program package},
	author       = {Sun, Qiming and Zhang, Xing and Banerjee, Samragni and Bao, Peng and Barbry, Marc and Blunt, Nick S. and Bogdanov, Nikolay A. and Booth, George H. and Chen, Jia and Cui, Zhi-Hao and Eriksen, Janus J. and Gao, Yang and Guo, Sheng and Hermann, Jan and Hermes, Matthew R. and Koh, Kevin and Koval, Peter and Lehtola, Susi and Li, Zhendong and Liu, Junzi and Mardirossian, Narbe and McClain, James D. and Motta, Mario and Mussard, Bastien and Pham, Hung Q. and Pulkin, Artem and Purwanto, Wirawan and Robinson, Paul J. and Ronca, Enrico and Sayfutyarova, Elvira R. and Scheurer, Maximilian and Schurkus, Henry F. and Smith, James E. T. and Sun, Chong and Sun, Shi-Ning and Upadhyay, Shiv and Wagner, Lucas K. and Wang, Xiao and White, Alec and Whitfield, James Daniel and Williamson, Mark J. and Wouters, Sebastian and Yang, Jun and Yu, Jason M. and Zhu, Tianyu and Berkelbach, Timothy C. and Sharma, Sandeep and Sokolov, Alexander Yu. and Chan, Garnet Kin-Lic},
	year         = 2020,
	month        = {07},
	journal      = {The Journal of Chemical Physics},
	volume       = 153,
	number       = 2,
	pages        = {024109},
	doi          = {10.1063/5.0006074}
}

@misc{Li:2024,
	title        = {Introducing {GPU}-acceleration into the {Python}-based Simulations of Chemistry Framework},
	author       = {Rui Li and Qiming Sun and Xing Zhang and Garnet Kin-Lic Chan},
	year         = 2024,
	url          = {https://arxiv.org/abs/2407.09700},
	eprint       = {2407.09700},
	archiveprefix = {arXiv},
	primaryclass = {physics.comp-ph}
}

@misc{Wu:2024,
	title        = {Enhancing {GPU}-acceleration in the {Python}-based Simulations of Chemistry Framework},
	author       = {Xiaojie Wu and Qiming Sun and Zhichen Pu and Tianze Zheng and Wenzhi Ma and Wen Yan and Xia Yu and Zhengxiao Wu and Mian Huo and Xiang Li and Weiluo Ren and Sheng Gong and Yumin Zhang and Weihao Gao},
	year         = 2024,
	url          = {https://arxiv.org/abs/2404.09452},
	eprint       = {2404.09452},
	archiveprefix = {arXiv},
	primaryclass = {physics.comp-ph}
}

@article{Whitten:1973,
	title        = {{Coulombic} potential energy integrals and approximations},
	author       = {Whitten, J. L.},
	year         = 1973,
	month        = {05},
	journal      = {The Journal of Chemical Physics},
	volume       = 58,
	number       = 10,
	pages        = {4496--4501},
	doi          = {10.1063/1.1679012}
}

@article{Weigend:2002,
	title        = {A fully direct {RI-HF} algorithm: Implementation, optimised auxiliary basis sets, demonstration of accuracy and efficiency},
	author       = {Weigend, Florian},
	year         = 2002,
	journal      = {Physical Chemistry Chemical Physics},
	publisher    = {The Royal Society of Chemistry},
	volume       = 4,
	pages        = {4285--4291},
	doi          = {10.1039/B204199P},
	issue        = 18
}

@article{Sodt:2006,
	title        = {Linear scaling density fitting},
	author       = {Sodt, Alex and Subotnik, Joseph E. and Head-Gordon, Martin},
	year         = 2006,
	month        = 11,
	journal      = {The Journal of Chemical Physics},
	volume       = 125,
	number       = 19,
	pages        = 194109,
	doi          = {10.1063/1.2370949}
}

@article{Aquilante:2007,
	title        = {Unbiased auxiliary basis sets for accurate two-electron integral approximations},
	author       = {Aquilante, Francesco and Lindh, Roland and Bondo Pedersen, Thomas},
	year         = 2007,
	month        = {09},
	journal      = {The Journal of Chemical Physics},
	volume       = 127,
	number       = 11,
	pages        = 114107,
	doi          = {10.1063/1.2777146}
}

@article{Aquilante:2009,
	title        = {Atomic {Cholesky} decompositions: A route to unbiased auxiliary basis sets for density fitting approximation with tunable accuracy and efficiency},
	author       = {Aquilante, Francesco and Gagliardi, Laura and Pedersen, Thomas Bondo and Lindh, Roland},
	year         = 2009,
	month        = {04},
	journal      = {The Journal of Chemical Physics},
	volume       = 130,
	number       = 15,
	pages        = 154107,
	doi          = {10.1063/1.3116784}
}

@article{Reine:2008,
	title        = {Variational and robust density fitting of four-center two-electron integrals in local metrics},
	author       = {Reine, Simen and Tellgren, Erik and Krapp, Andreas and Kjærgaard, Thomas and Helgaker, Trygve and Jansik, Branislav and Høst, Stinne and Salek, Paweł},
	year         = 2008,
	month        = {09},
	journal      = {The Journal of Chemical Physics},
	volume       = 129,
	number       = 10,
	pages        = 104101,
	doi          = {10.1063/1.2956507}
}

@article{Merlot:2013,
	title        = {Attractive electron–electron interactions within robust local fitting approximations},
	author       = {Merlot, Patrick and Kjærgaard, Thomas and Helgaker, Trygve and Lindh, Roland and Aquilante, Francesco and Reine, Simen and Pedersen, Thomas Bondo},
	year         = 2013,
	journal      = {Journal of Computational Chemistry},
	volume       = 34,
	number       = 17,
	pages        = {1486--1496},
	doi          = {10.1002/jcc.23284}
}

@article{Eichkorn:1995,
	title        = {Auxiliary basis sets to approximate {Coulomb} potentials},
	author       = {Karin Eichkorn and Oliver Treutler and Holger Öhm and Marco Häser and Reinhart Ahlrichs},
	year         = 1995,
	journal      = {Chemical Physics Letters},
	volume       = 240,
	number       = 4,
	pages        = {283--290},
	doi          = {10.1016/0009-2614(95)00621-A}
}

@article{Werner:2003,
	title        = {Fast linear scaling second-order Møller-Plesset perturbation theory (MP2) using local and density fitting approximations},
	author       = {Werner, Hans-Joachim and Manby, Frederick R. and Knowles, Peter J.},
	year         = 2003,
	month        = {05},
	journal      = {The Journal of Chemical Physics},
	volume       = 118,
	number       = 18,
	pages        = {8149--8160},
	doi          = {10.1063/1.1564816}
}

@article{Zhou:2006,
	title        = {Self-consistent-field calculations using Chebyshev-filtered subspace iteration},
	author       = {Yunkai Zhou and Yousef Saad and Murilo L. Tiago and James R. Chelikowsky},
	year         = 2006,
	journal      = {Journal of Computational Physics},
	volume       = 219,
	number       = 1,
	pages        = {172--184},
	doi          = {10.1016/j.jcp.2006.03.017}
}

@article{Lin:2016,
	title        = {Adaptively Compressed Exchange Operator},
	author       = {Lin, Lin},
	year         = 2016,
	journal      = {Journal of Chemical Theory and Computation},
	volume       = 12,
	number       = 5,
	pages        = {2242--2249},
	doi          = {10.1021/acs.jctc.6b00092}
}

@article{Hu:2017-ACE,
	title        = {Adaptively Compressed Exchange Operator for Large-Scale Hybrid Density Functional Calculations with Applications to the Adsorption of Water on Silicene},
	author       = {Hu, Wei and Lin, Lin and Banerjee, Amartya S. and Vecharynski, Eugene and Yang, Chao},
	year         = 2017,
	journal      = {Journal of Chemical Theory and Computation},
	volume       = 13,
	number       = 3,
	pages        = {1188--1198},
	doi          = {10.1021/acs.jctc.6b01184}
}

@article{Lu:2015,
	title        = {Compression of the electron repulsion integral tensor in tensor hypercontraction format with cubic scaling cost},
	author       = {Jianfeng Lu and Lexing Ying},
	year         = 2015,
	journal      = {Journal of Computational Physics},
	volume       = 302,
	pages        = {329--335},
	doi          = {10.1016/j.jcp.2015.09.014}
}

@article{Hu:2017-ISDF,
	title        = {Interpolative Separable Density Fitting Decomposition for Accelerating Hybrid Density Functional Calculations with Applications to Defects in Silicon},
	author       = {Hu, Wei and Lin, Lin and Yang, Chao},
	year         = 2017,
	journal      = {Journal of Chemical Theory and Computation},
	volume       = 13,
	number       = 11,
	pages        = {5420--5431},
	doi          = {10.1021/acs.jctc.7b00807}
}

@article{Dong:2018,
	title        = {Interpolative Separable Density Fitting through Centroidal Voronoi Tessellation with Applications to Hybrid Functional Electronic Structure Calculations},
	author       = {Dong, Kun and Hu, Wei and Lin, Lin},
	year         = 2018,
	journal      = {Journal of Chemical Theory and Computation},
	volume       = 14,
	number       = 3,
	pages        = {1311--1320},
	doi          = {10.1021/acs.jctc.7b01113}
}

@article{Olivares:2010,
	title        = {Accelerating Correlated Quantum Chemistry Calculations Using Graphical Processing Units and a Mixed Precision Matrix Multiplication Library},
	author       = {Olivares-Amaya, Roberto and Watson, Mark A. and Edgar, Richard G. and Vogt, Leslie and Shao, Yihan and Aspuru-Guzik, Alán},
	year         = 2010,
	journal      = {Journal of Chemical Theory and Computation},
	volume       = 6,
	number       = 1,
	pages        = {135--144},
	doi          = {10.1021/ct900543q}
}

@article{DePrince:2011,
	title        = {Coupled Cluster Theory on Graphics Processing Units {I.} The Coupled Cluster Doubles Method},
	author       = {DePrince, A. Eugene III and Hammond, Jeff R.},
	year         = 2011,
	journal      = {Journal of Chemical Theory and Computation},
	volume       = 7,
	number       = 5,
	pages        = {1287--1295},
	doi          = {10.1021/ct100584w}
}

@article{Pokhilko:2018,
	title        = {Double Precision Is Not Needed for Many-Body Calculations: Emergent Conventional Wisdom},
	author       = {Pokhilko, Pavel and Epifanovsky, Evgeny and Krylov, Anna I.},
	year         = 2018,
	journal      = {Journal of Chemical Theory and Computation},
	volume       = 14,
	number       = 8,
	pages        = {4088--4096},
	doi          = {10.1021/acs.jctc.8b00321}
}

@article{Huang:2020,
	title        = {Techniques for high-performance construction of {Fock} matrices},
	author       = {Huang, Hua and Sherrill, C. David and Chow, Edmond},
	year         = 2020,
	month        = {Jan},
	journal      = {The Journal of Chemical Physics},
	volume       = 152,
	number       = 2,
	pages        = {024122},
	doi          = {10.1063/1.5129452},
	issn         = {0021-9606, 1089-7690}
}

@inproceedings{Huang:2018,
	title        = {Accelerating Quantum Chemistry with Vectorized and Batched Integrals},
	author       = {Huang, Hua and Chow, Edmond},
	year         = 2018,
	month        = {Nov},
	booktitle    = {Proceedings of the International Conference for High Performance Computing, Networking, Storage and Analysis},
	publisher    = {IEEE},
	address      = {Dallas, TX, USA},
	pages        = {529--542},
	doi          = {10.1109/SC.2018.00044},
	isbn         = {978-1-5386-8384-2}
}

@inproceedings{Okuta:2017,
	title        = {{CuPy}: A {NumPy}-Compatible Library for {NVIDIA} {GPU} Calculations},
	author       = {Okuta, Ryosuke and Unno, Yuya and Nishino, Daisuke and Hido, Shohei and Loomis, Crissman},
	year         = 2017,
	booktitle    = {Proceedings of Workshop on Machine Learning Systems (LearningSys) in The Thirty-first Annual Conference on Neural Information Processing Systems (NIPS)}
}

@misc{CUTLASS:2023,
	title        = {{CUTLASS}},
	author       = {Thakkar, Vijay and Ramani, Pradeep and Cecka, Cris and Shivam, Aniket and Lu, Honghao and Yan, Ethan and Kosaian, Jack and Hoemmen, Mark and Wu, Haicheng and Kerr, Andrew and Nicely, Matt and Merrill, Duane and Blasig, Dustyn and Qiao, Fengqi and Majcher, Piotr and Springer, Paul and Hohnerbach, Markus and Wang, Jin and Gupta, Manish},
	year         = 2023,
	month        = 1,
	url          = {https://github.com/NVIDIA/cutlass}
}

@article{Krishnan:1980,
	title        = {Self‐consistent molecular orbital methods. {XX.} A basis set for correlated wave functions},
	author       = {Krishnan, R. and Binkley, J. S. and Seeger, R. and Pople, J. A.},
	year         = 1980,
	month        = {01},
	journal      = {The Journal of Chemical Physics},
	volume       = 72,
	number       = 1,
	pages        = {650--654},
	doi          = {10.1063/1.438955},
	issn         = {0021-9606}
}

@article{Weigend:2005,
	title        = {Balanced basis sets of split valence{,} triple zeta valence and quadruple zeta valence quality for {H} to {Rn}: Design and assessment of accuracy},
	author       = {Weigend, Florian and Ahlrichs, Reinhart},
	year         = 2005,
	journal      = {Physical Chemistry Chemical Physics},
	publisher    = {The Royal Society of Chemistry},
	volume       = 7,
	pages        = {3297--3305},
	doi          = {10.1039/B508541A},
	issue        = 18
}

@article{Weigend:2006,
	title        = {Accurate {Coulomb}-fitting basis sets for {H} to {Rn}},
	author       = {Weigend, Florian},
	year         = 2006,
	journal      = {Physical Chemistry Chemical Physics},
	publisher    = {The Royal Society of Chemistry},
	volume       = 8,
	pages        = {1057--1065},
	doi          = {10.1039/B515623H},
	issue        = 9
}

@article{Becke:1993,
	title        = {Density‐functional thermochemistry. {III.} The role of exact exchange},
	author       = {Becke, Axel D.},
	year         = 1993,
	month        = {04},
	journal      = {The Journal of Chemical Physics},
	volume       = 98,
	number       = 7,
	pages        = {5648--5652},
	doi          = {10.1063/1.464913},
	issn         = {0021-9606}
}

@article{Stephens:1994,
	title        = {{Ab Initio} Calculation of Vibrational Absorption and Circular Dichroism Spectra Using Density Functional Force Fields},
	author       = {Stephens, P. J. and Devlin, F. J. and Chabalowski, C. F. and Frisch, M. J.},
	year         = 1994,
	journal      = {The Journal of Physical Chemistry},
	volume       = 98,
	number       = 45,
	pages        = {11623--11627},
	doi          = {10.1021/j100096a001}
}

@article{NWChem:2020,
	title        = {NWChem: Past, present, and future},
	author       = {Aprà,E.  and Bylaska,E. J.  and de Jong,W. A.  and Govind,N.  and Kowalski,K.  and Straatsma,T. P.  and Valiev,M.  and van Dam,H. J. J.  and Alexeev,Y.  and Anchell,J.  and Anisimov,V.  and Aquino,F. W.  and Atta-Fynn,R.  and Autschbach,J.  and Bauman,N. P.  and Becca,J. C.  and Bernholdt,D. E.  and Bhaskaran-Nair,K.  and Bogatko,S.  and Borowski,P.  and Boschen,J.  and Brabec,J.  and Bruner,A.  and Cauët,E.  and Chen,Y.  and Chuev,G. N.  and Cramer,C. J.  and Daily,J.  and Deegan,M. J. O.  and Dunning,T. H.  and Dupuis,M.  and Dyall,K. G.  and Fann,G. I.  and Fischer,S. A.  and Fonari,A.  and Früchtl,H.  and Gagliardi,L.  and Garza,J.  and Gawande,N.  and Ghosh,S.  and Glaesemann,K.  and Götz,A. W.  and Hammond,J.  and Helms,V.  and Hermes,E. D.  and Hirao,K.  and Hirata,S.  and Jacquelin,M.  and Jensen,L.  and Johnson,B. G.  and Jónsson,H.  and Kendall,R. A.  and Klemm,M.  and Kobayashi,R.  and Konkov,V.  and Krishnamoorthy,S.  and Krishnan,M.  and Lin,Z.  and Lins,R. D.  and Littlefield,R. J.  and Logsdail,A. J.  and Lopata,K.  and Ma,W.  and Marenich,A. V.  and Martin del Campo,J.  and Mejia-Rodriguez,D.  and Moore,J. E.  and Mullin,J. M.  and Nakajima,T.  and Nascimento,D. R.  and Nichols,J. A.  and Nichols,P. J.  and Nieplocha,J.  and Otero-de-la-Roza,A.  and Palmer,B.  and Panyala,A.  and Pirojsirikul,T.  and Peng,B.  and Peverati,R.  and Pittner,J.  and Pollack,L.  and Richard,R. M.  and Sadayappan,P.  and Schatz,G. C.  and Shelton,W. A.  and Silverstein,D. W.  and Smith,D. M. A.  and Soares,T. A.  and Song,D.  and Swart,M.  and Taylor,H. L.  and Thomas,G. S.  and Tipparaju,V.  and Truhlar,D. G.  and Tsemekhman,K.  and Van Voorhis,T.  and Vázquez-Mayagoitia,Á.  and Verma,P.  and Villa,O.  and Vishnu,A.  and Vogiatzis,K. D.  and Wang,D.  and Weare,J. H.  and Williamson,M. J.  and Windus,T. L.  and Woliński,K.  and Wong,A. T.  and Wu,Q.  and Yang,C.  and Yu,Q.  and Zacharias,M.  and Zhang,Z.  and Zhao,Y.  and Harrison,R. J.},
	year         = 2020,
	journal      = {The Journal of Chemical Physics},
	volume       = 152,
	number       = 18,
	pages        = 184102,
	doi          = {10.1063/5.0004997}
}

@article{Psi4:2020,
	title        = {{PSI4} 1.4: Open-source software for high-throughput quantum chemistry},
	author       = {Smith, Daniel G. A. and Burns, Lori A. and Simmonett, Andrew C. and Parrish, Robert M. and Schieber, Matthew C. and Galvelis, Raimondas and Kraus, Peter and Kruse, Holger and Di Remigio, Roberto and Alenaizan, Asem and James, Andrew M. and Lehtola, Susi and Misiewicz, Jonathon P. and Scheurer, Maximilian and Shaw, Robert A. and Schriber, Jeffrey B. and Xie, Yi and Glick, Zachary L. and Sirianni, Dominic A. and O’Brien, Joseph Senan and Waldrop, Jonathan M. and Kumar, Ashutosh and Hohenstein, Edward G. and Pritchard, Benjamin P. and Brooks, Bernard R. and Schaefer, Henry F., III and Sokolov, Alexander Yu. and Patkowski, Konrad and DePrince, A. Eugene, III and Bozkaya, Uğur and King, Rollin A. and Evangelista, Francesco A. and Turney, Justin M. and Crawford, T. Daniel and Sherrill, C. David},
	year         = 2020,
	month        = {05},
	journal      = {The Journal of Chemical Physics},
	volume       = 152,
	number       = 18,
	pages        = 184108,
	doi          = {10.1063/5.0006002}
}

@article{ORCA:2020,
	title        = {The {ORCA} quantum chemistry program package},
	author       = {Neese, Frank and Wennmohs, Frank and Becker, Ute and Riplinger, Christoph},
	year         = 2020,
	month        = {06},
	journal      = {The Journal of Chemical Physics},
	volume       = 152,
	number       = 22,
	pages        = 224108,
	doi          = {10.1063/5.0004608}
}

@article{Lehtola:2018,
	title        = {Recent developments in libxc - A comprehensive library of functionals for density functional theory},
	author       = {Susi Lehtola and Conrad Steigemann and Micael J.T. Oliveira and Miguel A.L. Marques},
	year         = 2018,
	journal      = {SoftwareX},
	volume       = 7,
	pages        = {1--5},
	doi          = {10.1016/j.softx.2017.11.002},
	issn         = {2352-7110}
}

@article{Ozaki:2012,
	title        = {Error-free transformations of matrix multiplication by using fast routines of matrix multiplication and its applications},
	author       = {Ozaki, Katsuhisa and Ogita, Takeshi and Oishi, Shin'Ichi and Rump, Siegfried M.},
	year         = 2012,
	month        = jan,
	journal      = {Numer. Algorithms},
	publisher    = {Springer-Verlag},
	address      = {Berlin, Heidelberg},
	volume       = 59,
	number       = 1,
	pages        = {95–118},
	doi          = {10.1007/s11075-011-9478-1},
	issn         = {1017-1398},
	numpages     = 24
}

@article{Ozaki:2013,
    title        = {Generalization of error-free transformation for matrix multiplication and its application},
    author       = {Katsuhisa Ozaki and Takeshi Ogita and Shin'ichi Oishi and Siegfried M. Rump},
    journal      = {Nonlinear Theory and Its Applications, IEICE},
    volume       = {4},
    number       = {1},
    pages        = {2-11},
    year         = {2013},
    doi          = {10.1587/nolta.4.2}
}

@inproceedings{Mukunoki:2020,
	title        = {{DGEMM} Using {Tensor Cores}, and Its Accurate and Reproducible Versions},
	author       = {Mukunoki, Daichi and Ozaki, Katsuhisa and Ogita, Takeshi and Imamura, Toshiyuki},
	editor       = {Sadayappan, Ponnuswamy and Chamberlain, Bradford L. and Juckeland, Guido and Ltaief, Hatem},
	year         = 2020,
	booktitle    = {High Performance Computing},
	publisher    = {Springer International Publishing},
	pages        = {230--248},
	doi          = {10.1007/978-3-030-50743-5_12},
	isbn         = {978-3-030-50743-5}
}

@article{Ootomo:2022,
	title        = {Recovering single precision accuracy from {Tensor Cores} while surpassing the {FP32} theoretical peak performance},
	author       = {Hiroyuki Ootomo and Rio Yokota},
	year         = 2022,
	journal      = {The International Journal of High Performance Computing Applications},
	volume       = 36,
	number       = 4,
	pages        = {475--491},
	doi          = {10.1177/10943420221090256}
}

@article{Ootomo:2024,
	title        = {{DGEMM} on integer matrix multiplication unit},
	author       = {Hiroyuki Ootomo and Katsuhisa Ozaki and Rio Yokota},
	year         = 2024,
	journal      = {The International Journal of High Performance Computing Applications},
	volume       = 38,
	number       = 4,
	pages        = {297--313},
	doi          = {10.1177/10943420241239588}
}

@misc{Ootomo:ozIMMU,
	title        = {{ozIMMU} - {DGEMM} on Int8 {Tensor Core}},
	author       = {Hiroyuki Ootomo},
	year         = {2025},
	publisher    = {GitHub},
	note         = {Accessed: 2025-07-29},
	howpublished = {\url{https://github.com/enp1s0/ozIMMU}}
}

@misc{Ozaki:2025,
	title        = {Ozaki Scheme {II}: A {GEMM}-oriented emulation of floating-point matrix multiplication using an integer modular technique},
	author       = {Katsuhisa Ozaki and Yuki Uchino and Toshiyuki Imamura},
	year         = 2025,
	url          = {https://arxiv.org/abs/2504.08009},
	eprint       = {2504.08009},
}

@misc{Ozaki:GEMMul8,
	title        = {{GEMMul8} - {GEMM} emulation using int8 matrix engines based on the {Ozaki Scheme II}},
	author       = {Katsuhisa Ozaki and Yuki Uchino and Toshiyuki Imamura},
	year         = {2025},
	publisher    = {GitHub},
	note         = {Accessed: 2025-07-31},
	howpublished = {\url{https://github.com/RIKEN-RCCS/GEMMul8}}
}

@inproceedings{Schwarz:2026,
    author       = {Schwarz, Angelika and Anders, Anton and Brower, Cole and Bayraktar, Harun and Gunnels, John and Clark, Kate and Xu, RuQing G. and Rodriguez, Samuel and Cayrols, Sebastien and Tabaszewski, Pawel and Podlozhnyuk, Victor},
    title        = {Guaranteed {DGEMM} Accuracy While Using Reduced Precision {Tensor Cores} Through Extensions of the {Ozaki Scheme}},
    year         = {2026},
    isbn         = {9798400720673},
    publisher    = {Association for Computing Machinery},
    address      = {New York, NY, USA},
    doi          = {10.1145/3773656.3773670},
    booktitle    = {Proceedings of the Supercomputing Asia and International Conference on High Performance Computing in Asia Pacific Region},
    pages        = {91–101},
    numpages     = {11},
    series       = {SCA/HPCAsia '26}
}

@misc{cuEST:2026,
	title        = {{cuEST}: Accelerating Quantum Chemistry on {NVIDIA} {GPU}s},
	author       = {NVIDIA},
	year         = {2026},
	urldate      = {2026-03-20},
	howpublished = {\url{https://docs.nvidia.com/cuda/cuest/}}
}

@inproceedings{Thall2006ExtendedPrecision,
  author    = {Andrew Thall},
  title     = {Extended-Precision Floating-Point Numbers for {GPU} Computation},
  booktitle = {{ACM} {SIGGRAPH} 2006 Research Posters},
  year      = {2006},
  publisher = {ACM},
  address   = {New York, NY, USA},
  pages     = {52--es},
  doi       = {10.1145/1179622.1179682},
  url       = {https://dl.acm.org/doi/10.1145/1179622.1179682}
}

@article{Vydrov2010JCP,
  author  = {Vydrov, Oleg A. and Van Voorhis, Troy},
  title   = {Nonlocal {van der Waals} density functional: The simpler the better},
  journal = {The Journal of Chemical Physics},
  year    = {2010},
  volume  = {133},
  number  = {24},
  pages   = {244103},
  doi     = {10.1063/1.3521275}
}

@article{Dawson2024RedPrecisionQC,
  author  = {Dawson, William and Ozaki, Katsuhisa and Domke, Jens and Nakajima, Takahito},
  title   = {Reducing Numerical Precision Requirements in Quantum Chemistry Calculations},
  journal = {Journal of Chemical Theory and Computation},
  year    = {2024},
  volume  = {20},
  number  = {24},
  pages   = {10826-10837},
  doi     = {10.1021/acs.jctc.4c00938},
  url     = {https://doi.org/10.1021/acs.jctc.4c00938}
}

\newpage

\begin{figure}[]
\centering
\includegraphics[width=\linewidth]{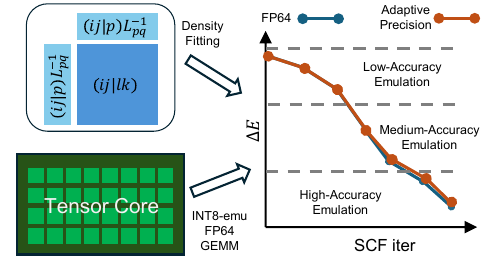}  
\caption{TOC Graphic.}
\label{fig:toc}
\end{figure}



  
  

\end{document}